\documentclass[10pt,journal,final]{IEEEtran}
\IEEEoverridecommandlockouts

\usepackage{amsfonts,amssymb}
\usepackage{ifpdf}
\usepackage{cite}
\usepackage{stfloats}
\usepackage{bm}
\usepackage{color,xcolor}
\usepackage{subfigure}
\ifCLASSINFOpdf
\usepackage[pdftex]{graphicx}
\graphicspath{{../pdf/}{../jpeg/}}
\DeclareGraphicsExtensions{.pdf,.jpeg,.png}
\else
\usepackage[dvips]{graphicx}

\graphicspath{{../eps/}}

\DeclareGraphicsExtensions{.eps}
\fi

\usepackage[cmex10]{amsmath}

\usepackage{algorithm}
\usepackage{algorithmic}
\ifCLASSOPTIONcompsoc
\else
\usepackage[caption=false,font=footnotesize]{subfig}
\fi

\usepackage{makecell}

\begin{document}	
%\title{DeepJSCC-Enabled Image Transmission Over MIMO systems Via LLM-enabled In-Context Learning}
%\title{LLM-Enabled In-Context Learning for DeepJSCC of Image Transmission over MIMO systems}
\title{In-Context Learning for Deep Joint Source-Channel Coding Over MIMO Channels}
\author{
	Meng Hua,~\IEEEmembership{Senior Member,~IEEE,}
	Wenjing Zhang,~\IEEEmembership{Student Member,~IEEE,}
	Chenghong~Bian,~\IEEEmembership{Student Member,~IEEE},
	and Deniz~G\"und\"uz,~\IEEEmembership{Fellow,~IEEE}
	
	\thanks{M. Hua, C. Bian, and D. G\"und\"uz are with the Department of Electrical and Electronic Engineering, Imperial College London, London SW7 2AZ, U.K. (e-mail: \{m.hua, c.bian22, d.gunduz\}@imperial.ac.uk); W. Zhang is with the Beijing Laboratory of Advanced Information Network, Beijing University of Posts and Telecommunications, Beijing 100876, China (e-mail: zhangwenjing@bupt.edu.cn).
	}
	
}
\maketitle
\vspace{-1.2cm}
\begin{abstract}	
Large language models have demonstrated the ability to perform \textit{in-context learning} (ICL), whereby the model performs predictions by directly mapping the query and a few examples from the given task to the output variable. In this paper, we study  ICL for deep joint source-channel coding (DeepJSCC) in image transmission over multiple-input multiple-output (MIMO) systems, where an ICL denoiser is employed for  MIMO symbol estimation. We first study the transceiver without any hardware impairments and explore the integration of transformer-based ICL with DeepJSCC in both open-loop and closed-loop MIMO systems, depending on the availability of channel state information (CSI) at the transceiver. For both open-loop and closed-loop scenarios,  we propose two MIMO transceiver architectures that leverage context information, i.e., pilot sequences and their outputs, as additional inputs, enabling the DeepJSCC encoder, DeepJSCC decoder, and the ICL  denoiser to jointly learn encoding, decoding, and estimation strategies tailored to each channel realization. Next, we extend our study to a more challenging scenario where the transceiver suffers from in-phase and quadrature (IQ) imbalance, resulting in nonlinear MIMO estimation. In this case, the context information is also exploited, facilitating joint learning across the DeepJSCC encoder, decoder, and the ICL  denoiser under hardware impairments and varying channel conditions. Numerical results demonstrate that the ICL  denoiser for MIMO estimation significantly outperforms the conventional least-squares method, with even greater advantages under IQ imbalance. Moreover, the proposed transformer-based ICL framework, integrated with contextual information, achieves significant improvements in end-to-end image reconstruction quality under transceiver IQ imbalance.
\end{abstract}
\begin{IEEEkeywords}
Deep joint source-channel coding (DeepJSCC),  in-context learning (ICL),  in-phase and quadrature (IQ)  imbalance, non-linear estimation.
\end{IEEEkeywords}

\section{Introduction}
The upcoming sixth-generation (6G) wireless networks are expected to support a wide range of emerging applications, such as immersive extended reality \cite{shen2023toward} and autonomous driving \cite{yang2021edge}, all of which demand ultra-reliable and low-latency communication. Deep joint source-channel coding (DeepJSCC) has been proposed as a promising solution to support these applications \cite{Bian2025process,xu2022wirless,kurka2021bandwidth,yang2025swinjscc,yang2022ofdm,Wu2022channel}. DeepJSCC is a deep learning-based communication framework that jointly performs source and channel coding using neural networks, rather than treating them as two separate processes.  In particular, this approach eliminates the need for explicit source compression and error correction, instead relying on a unified encoder parameterized by neural networks. Unlike conventional separated designs, DeepJSCC enables end-to-end optimization and exhibits significant performance gains in the practical finite-block length regime \cite{Bourtsoulatze2019deep} and \cite{joint2024gunduz}.

There have been extensive research activities  on DeepJSCC for image and video transmission over wireless channels. The seminal work in \cite{Bourtsoulatze2019deep} proposed a convolutional neural network (CNN)-based end-to-end JSCC architecture for image transmission, demonstrating that DeepJSCC can outperform conventional separate schemes, especially by avoiding the so-called ``cliff effect", where the performance sharply degrades once the channel conditions fall below a certain threshold.
Subsequently, \cite{wu2024transformer} introduced a vision transformer (ViT)-based end-to-end DeepJSCC architecture, leveraging global self-attention mechanisms for a more discriminative semantic feature representation compared to CNN-based approaches. Then, this approach was extended to multiple-input multiple-output (MIMO) systems \cite{bian2023space,wu2024deepmimo,xu2023deep,Inokuma2024performance,jiang2024deep}. 
%Although the aforementioned works have demonstrated impressive results of DeepJSCC, they often assume that the channel state information (CSI) is perfectly known at either the transmitter, the receiver, or both. However, in practical communication scenarios, perfect CSI is not always available, especially in MIMO systems. 
%Due to the mismatch between the model's assumptions and the actual channel conditions, symbol estimation at the receiver under the perfect CSI assumption can result in significant estimation errors. 
In addition, a handful of works, e.g., \cite{xu2022wirless,Wu2022channel,wu2024deepmimo,jiang2024deep}, have demonstrated that incorporating channel state information (CSI)-related features into DeepJSCC architectures via attention mechanisms can significantly enhance system adaptability to dynamic channel conditions by adjusting feature importance. 
Nevertheless, the aforementioned works assume perfect knowledge of CSI, which is overly idealistic and impractical in real-world scenarios.

To estimate transmit symbols over fading channels, traditional methods typically adopt a two-step approach: first, estimating the channel based on the pilots and using techniques such as the least squares (LS) estimator, and then detecting symbols based on the estimated channel parameters. However, this two-step approach suffers from double estimation errors and also involves intensive computation due to matrix inversion operations. To overcome these limitations, data-driven approaches using deep learning have emerged as promising alternatives. Deep learning-based methods can directly learn channel estimation and symbol detection, potentially improving robustness and efficiency. Deep learning methods, such as recurrent neural networks (RNNs) \cite{lu2019mimo} and CNNs \cite{neumann2018learning}, have been explored for this task. Nevertheless, adapting pre-trained DNNs to new wireless environments often results in significant performance loss, and the challenge of maintaining performance with few samples remains unresolved \cite{he2020model,ye2018power,Soltani2019deep}.

Recently, large pre-trained sequence models, also known as large language models (LLMs), such as GPT \cite{radford2018improving}, have emerged as an alternative, owing to their ability to perform \textit{in-context learning} (ICL) \cite{zhou2024large,abbas2024leveraging,zhou2025generative}. ICL refers to a model's ability to learn and adapt to a task based on the information provided in the prompt, without any changes to its underlying parameters \cite{dong2022survey,brown2020language,garg2022can}. The model learns from the examples given in the prompt and predicts the next output, even if it has not been explicitly trained to memorize that specific example. The core principle of ICL is to leverage the transformer architecture, where the model uses attention mechanisms to focus on relevant parts of the prompt and generalize from them during inference. Particularly,  \cite{garg2022can} was the first to experimentally study linear regression via ICL, verifying that the model can predict  $ f(x_{\text{query}})$ with a given prompt sequence $\left(x_1, f(x_1), \ldots, x_k, f(x_k), x_{\text{query}}\right)$, where $f\left(  \cdot  \right)$ belongs to a class of linear functions. Inspired by the experimental study of ICL in \cite{garg2022can}, subsequent works have sought to develop a theoretical understanding of this phenomenon \cite{zhang2024trained,wu2023many}.  ICL effectively bundles the conventional two-step estimation into a single step, learning to decode the signal directly from the given context samples without incurring double estimation errors.
Subsequently, \cite{kunde2023transformers} studied a reverse problem, where the prompt $\left( {f({x_1}),{x_1}, \ldots ,f({x_k}),{x_k},f({x_{{\rm{query}}}})} \right)$ is used to predict input $x_{\rm {query}}$,  which differs from the linear regression problem in the presence of noisy observations, and theoretically proved that a single-layer softmax attention transformer is the minimizer
of the corresponding training loss.

In addition to these theoretical advances, several recent works have applied the ICL paradigm to wireless communication problems, where symbol detection at each channel signal-to-noise ratios (SNR) is treated as a distinct task, enabling the model to exploit task diversity for improved generalization.
Work \cite{zecchin2024incontext}  implemented equalization over non-linear MIMO channels, where  transmit symbols are recovered via ICL. This work was subsequently extended to the cell-free multi-user scenario in \cite{Zecchin2024cellfree} and \cite{song2024context}. To reduce the length of pilot sequences required for ICL-based equalization, a decision feedback in-context symbol detection over block-fading channels was proposed in  \cite{fan2024decision}, where previously detected symbols are iteratively incorporated into the context as additional inputs to enhance the estimation of subsequent symbols, thereby improving performance even with limited pilot data.

Inspired by the advantages of transformer-based ICL, in this paper, we integrate the ICL  denoiser into DeepJSCC. Thus, the conventional MIMO detector at the receiver can be replaced by the ICL denoiser. In addition, compared to existing works \cite{zecchin2024incontext,Zecchin2024cellfree,song2024context,fan2024decision}, where ICL is used only for symbol estimation, we further consider the context information, i.e., pilot sequences and their outputs, as additional inputs to both the DeepJSCC encoder and decoder. This design enables the DeepJSCC encoder,  decoder, and the ICL denoiser to simultaneously learn how to encode, decode, and detect for each specific channel realization. Moreover, previous works \cite{zecchin2024incontext,Zecchin2024cellfree,song2024context,fan2024decision} assume perfect transceiver hardware, whereas practical hardware impairments, such as in-phase and quadrature (IQ)  imbalance \cite{soleymani2022improper,javed2019multiple,Soleymani2020improperx}, are not considered. In fact, due to IQ imbalance, symbol estimation becomes a non-linear estimation problem, which poses a significantly more challenging task. In this paper, we explore and demonstrate the potential of  ICL for non-linear MIMO symbol estimation.
The contributions of this paper are summarized as follows:
\begin{itemize}
	\item  We propose a framework for transformer-based ICL for DeepJSCC of image transmission over MIMO systems.  Particularly, we consider two MIMO setups, namely, open-loop and closed-loop systems. For open-loop MIMO systems, we assume that the context information, i.e., pilot sequences and their outputs, is only available at the receiver, while for closed-loop MIMO systems, we assume that the context information is available at both the transmitter and receiver. 
	\item  For open-loop MIMO systems, we design two different symbol estimation paradigms via ICL. In the first paradigm,  transmit symbols are directly estimated by the ICL  denoiser. In the second paradigm, transmit symbols pass through an inverted MIMO channel and are then estimated via ICL. For both estimation paradigms, additional context information including channel heatmap or implicit channel-aware representation (ICAR) is  provided to the DeepJSCC decoder. 
	\item For closed-loop MIMO systems, we also design two different symbol estimation paradigms via ICL. The first paradigm is similar to the first one for open-loop systems, but the channel heatmap is sent to both the DeepJSCC encoder and decoder. In the second paradigm, a singular value decomposition (SVD)-based channel for symbol estimation via ICL is proposed, with the channel heatmap or ICAR similarly sent to both the DeepJSCC encoder and decoder.
	\item Next, we study transformer-based ICL for DeepJSCC  under transceiver IQ imbalance. Compared to previous works assuming perfect IQ balance, we propose two different symbol estimation paradigms for each MIMO setup, explicitly incorporating IQ imbalance into the ICL-based estimation process.
	\item Extensive numerical results verify the effectiveness of the ICL-based estimation method for both the IQ balance and IQ imbalance cases, and unveil the superiority of the proposed framework, showcasing significant improvements in image reconstruction quality compared to conventional LS-based channel equalization schemes.
\end{itemize} 
The rest of this paper is organized as follows. Section II introduces the system model, including the DeepJSCC framework and an ICL denoiser. Section III analyzes the transformer-based ICL for DeepJSCC under perfect IQ balance conditions, while Section IV extends this analysis to scenarios with IQ imbalance. Numerical results are presented in Section V, and finally, Section VI concludes the paper.

\emph{Notations}: Boldface upper-case and lower-case letters denote matrices and vectors, respectively. ${{\bf{1}}_{L}}$ represents a vector of  all ones  with the length of $L$.  ${\mathbb C}^ {d_1\times d_2}$ $({\mathbb R}^ {d_1\times d_2})$ stands for the set of  complex (real) $d_1\times d_2$  matrices.  ${\left\| {\bf{X}} \right\|_F}$  represents the  Frobenius norm of ${\bf{X}}$.
A circularly symmetric complex Gaussian (CSCG) scalar variable $ x$ with mean $  {\mu}$ and  covariance matrix ${\sigma ^2}$ is denoted by ${ x} \sim {\cal CN}\left( {{{ \mu }},{{\sigma^2 }}} \right)$. ${\left(  \cdot  \right)^*}$, ${\left(  \cdot  \right)^H}$,  ${\left(  \cdot  \right)^{\dag}}$,  ${\left(  \cdot  \right)^{-1}}$, and $ \odot $  stand for the conjugate operator,  conjugate transpose operator,  pseudo inverse operator, inverse operator, and Hadamard product operator, respectively.

\section{System Model}

\begin{figure}[!t]
	\centerline{\includegraphics[width=3.5in]{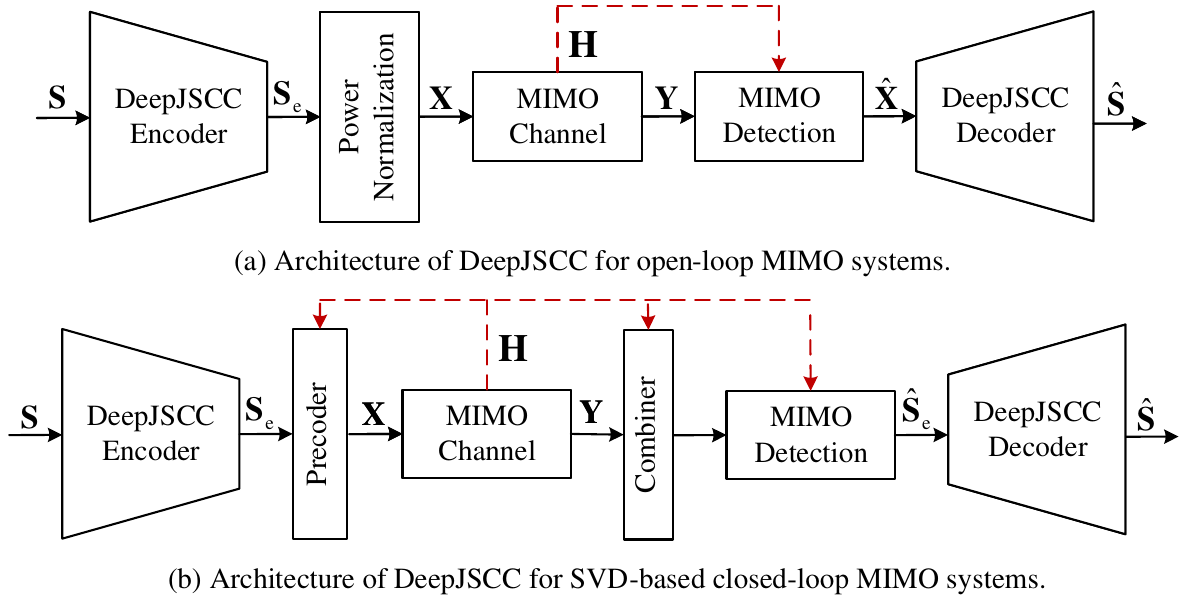}}
	\caption{Architectures of DeepJSCC for MIMO  systems.} \label{DeepJSCC}
	%	\vspace{-0.3cm}
\end{figure}

\subsection{DeepJSCC Model}
In this subsection, we discuss two types of DeepJSCC models based on whether CSI is available at the transmitter and/or the receiver. If CSI is  available only at the receiver, it corresponds to an \textit{open-loop} MIMO system, while  if it is available at both the transmitter and receiver, it corresponds to a \textit{closed-loop} MIMO system.

\subsubsection{Open-Loop MIMO Systems}
As illustrated in Fig.~\ref{DeepJSCC}(a), we consider an image reconstruction  task implemented  by DeepJSCC over an $M \times M$ MIMO system, where $M$ represents the number of transmit and receive antennas. Although images are used as the input modality in this example, the underlying formulation and results readily extend to other tasks and data types.
Let ${\mathbf{S}} \in {{\mathbb R}^{C \times H \times W}}$ denote the input image, where $C$, $H$, and $W$ represent the number of color channels, height, and width of the image, respectively. The transmitter encodes the input image ${\mathbf{S}}$ into a latent feature matrix denoted by ${{\mathbf{S}}_{\text{e}}} \in {{\mathbb R}^{p^2 \times \frac{{2ML}}{{{p^2}}}}}$ via an encoder, where ${{p^2}}$ denotes the number of patches of one image and $L$ represents the number of channel uses. Denoting by  ${f_{\bm \theta} }$ the encoding function parameterized by DNN parameters ${\bm \theta} $, we have 
\begin{align}
{{\bf{S}}_{\rm{e}}} = {f_{\bm \theta} }\left( {\bf{S}} \right). \label{open_loop_encoding}
\end{align}
Then, by converting ${{\mathbf{S}}_{\text{e}}}$ into a complex matrix ${\mathbf{X}} \in {{\mathbb C}^{M \times L}}$  for transmission over the air, we impose a transmit power constraint: 
\begin{align}
\frac{1}{{ML}}\left\| {\bf{X}} \right\|_F^2 \le P, \label{powerbudget}
\end{align}
where $P$ denotes the transmit power budget. 

We consider quasi-static block fading channels, assuming that the CSI remains constant during  each block, where a block corresponds to one image transmission, but may vary independently from one block to the next. Let ${\bf{H}} \in {{\mathbb C}^{{M} \times {M}}}$ denote the complex equivalent baseband  MIMO channel from the transmitter to the receiver. Then, the signal received at the receiver during one block of $L$ channel uses is given by
\begin{align}
{\bf{Y}} = {\bf{HX}} + {\bf{W}}, \label{MIMO_equation}
\end{align}
where ${\bf{W}} \in {{\mathbb C}^{M \times L}}$  denotes additive white Gaussian noise, with each entry in ${\bf{W}}$ independently distributed as ${{\bf{W}}_{i,j}} \sim {\cal CN}\left( {0,{\sigma ^2}} \right)$.

If the state of ${\bf{H}}$ in each block is perfectly known at the receiver, standard estimation techniques, such as LS or minimum mean square error (MMSE) estimators, can be applied to recover the transmitted symbols, denoted by ${{\bf{\hat X}}}$. Then, feeding ${{\bf{\hat X}}}$ into the DeepJSCC decoder, the output reconstructed image  ${{\bf{\hat S}}}$ can be obtained as 
\begin{align}
{\bf{\hat S}} = {g_{\bm \phi} }\left( {{\bf{\hat X}},{\bf{ H}},{\sigma^2}} \right), \label{decoderfunction}
\end{align}
where ${g_{\bm \phi} }\left(  \cdot  \right)$ denotes the decoding function parameterized by DNN parameters ${\bm \phi} $.

\subsubsection{SVD-based Closed-Loop MIMO  Systems}
For closed-loop MIMO systems,  CSI is available at both the DeepJSCC encoder and decoder. In contrast to  the encoding function given in \eqref{open_loop_encoding} for the open-loop systems, the encoding function  for the closed-loop  systems can incorporate CSI, and is denoted as 
\begin{align}
{{\bf{S}}_{\rm{e}}} = {f_{\bm \theta} }\left( {{\bf S},{\bf H},\sigma^2 } \right).
\end{align}
The transmitter can perform precoding to optimize  power allocation across data streams based on the available CSI. As shown in Fig.~\ref{DeepJSCC}(b),  given  channel ${\bf{H}} $, we perform   SVD of  ${\bf{H}} $  as  ${\bf{H}} = {\bf{U\Sigma }}{{\bf{V}}^H}$, where ${\bf{U}} \in {{\mathbb C}^{{M} \times {M}}}$ and ${\bf{V}} \in {{\mathbb C}^{{M} \times {M}}}$ are complex unitary matrices, and ${\bf{\Sigma}} \in {{\mathbb R}^{{M} \times {M}}}$ is a diagonal matrix with non-negative real entries. Based on the  SVD of channel ${{\bf{ H}}}$, the precoder and combiner are set to
${\bf{V}}$ and ${{\bf{U}}^H}$, respectively. To detect ${{\bf{S}}_{\rm{e}}}$, the received signal after combining is multiplied by ${{\bf{\Sigma }}^\dag }$, yielding
\begin{align}
{{{\bf{\hat S}}}_{\rm{e}}} &= {{\bf{\Sigma }}^\dag }{{\bf{U}}^H}\left( {{\bf{HV}}{{\bf{S}}_{\rm{e}}} + {\bf{W}}} \right)\notag\\
&\overset{(a)}{=}  {{\bf{S}}_{\rm{e}}} + {{\bf{\Sigma }}^\dag }{{\bf{U}}^H}{\bf{W}}, \label{closed_loop_detection}
\end{align}
where (a) holds if ${\bf{H}}$ is a full-rank matrix.

Finally, ${{{\bf{\hat S}}}_{\rm{e}}}$, together with CSI and noise information, is fed into the DeepJSCC decoder to recover the image as
\begin{align}
{\bf{\hat S}} = {g_{\bm \phi} }\left( {{{\bf{\hat S}}}_{\rm{e}},{\bf{ H}},{\sigma^2}} \right). 
\end{align}
\begin{figure}[!t]
	\centerline{\includegraphics[width=2.2in]{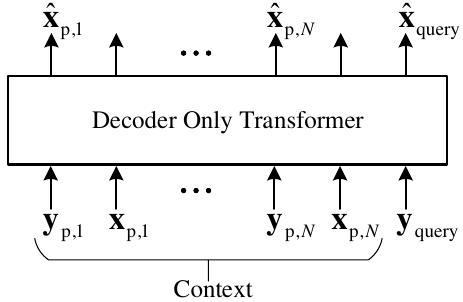}}
	\caption{Transformer-based architecture for ICL-based transmit symbol estimation.} \label{ICLmodel}
%	\vspace{-0.3cm}
\end{figure}
\subsection{ICL Denoiser} \label{ICL_architecture_model}
In traditional MIMO estimation methods, the receiver first estimates the channel using pilot sequences, and then performs symbol estimation based on the received symbol and the estimated channel to recover the transmitted symbol. However, this conventional two-step approach treats channel estimation and symbol estimation  separately, resulting in cumulative estimation errors and suboptimal performance, especially in very noisy environments.

Instead of using conventional estimation methods, we adopt an ICL denoiser to recover the channel input directly by removing the noise. 
 As demonstrated in \cite{garg2022can}, the decoder-only transformer architecture has shown the capability to perform ICL for inverse linear models: given a prompt comprising context information and a query output, the decoder-only transformer can infer the corresponding input.
As illustrated in Fig.~\ref{ICLmodel}, we use a decoder-only transformer of  GPT-2 as the backbone of  ICL  estimator in estimation tasks \cite{zecchin2024incontext}. Transformer  models have proven effective for many wireless tasks \cite{wang2022transformer} due to their ability to capture long-range dependencies and enhance generalization to diverse channel conditions. The provable optimality of transformers as in-context estimators 
 for wireless  systems is shown in \cite{kunde2023transformers}. By employing a masked self-attention mechanism, the model generates the predicted symbol ${{{\bf{\hat x}}}_{{\rm{query}}}}$ (i.e., estimated transmit symbol $ \bf \hat X$ in the open-loop system) at the same position as the observed symbol ${{{\bf{y}}_{{\rm{query}}}}}$ (i.e., received symbol $\bf Y$ in the open-loop system), relying solely on known prior context and the received signal. During inference, the transformer addresses a series of  estimation problems  for a given task, utilizing pilot sequences and their outputs as context. For example,  the authors in \cite{zecchin2024incontext} applied a decoder-only transformer  for MIMO detection, where the discrete transmit symbol is detected based on the context.
 
Before providing the formulation of the ICL-based symbol estimation problem, we first introduce some definitions for clarity.

\textbf{(1) Channel equalization task:}  Each ICL estimation task $\tau$ is represented by a tuple $\tau  = \left( {{\bf{H}},{\sigma ^2}} \right)$ consisting of CSI $\bf H$ and channel noise variance $\sigma ^2$. Note that the ICL  denoiser  has no prior knowledge of task $\tau$ and operates solely based on a prompt.

\textbf{(2) Context information:} Let ${{\bf{x}}_{{\rm{p}},n}} \in {{\mathbb C}^{M \times 1}}$ and ${{\bf{y}}_{{\rm{p}},n}} \in {{\mathbb C}^{M \times 1}}$ denote the $n$-th  transmitted and received pilot sequences, respectively,  satisfying
\begin{align}
 {{\bf{y}}_{{\rm{p}},n}} = {\bf{H}}{{\bf{x}}_{{\rm{p}},n}} + {\bf{w}}_{{\rm p},n},
 \end{align}
 as similarly defined in \eqref{MIMO_equation}.
 
Then, the context information for task $\tau$ is denoted by
\begin{align}
{C_{{\rm{p}}}^{\tau}} = \left\{ {\left( {{{\bf{y}}_{{\rm{p}},1}},{{\bf{x}}_{{\rm{p}},1}}} \right), \ldots ,\left( {{{\bf{y}}_{{\rm{p}},N}},{{\bf{x}}_{{\rm{p}},N}}} \right)} \right\},
\end{align} 
where $N$ is the pilot sequence length. Note that each pilot pair $\left( {{{\bf{y}}_{{\rm{p}},n}},{{\bf{x}}_{{\rm{p}},n}}} \right),\forall n,$ is independently and identically distributed  given task $\tau $.

\textbf{(3) Prompt:} A prompt for  task $\tau$ is defined as 
\begin{align}
{\rm prompt}=\left( {C_{\rm{p}}^\tau ,{{\bf y}_{{\rm{query}}}}} \right),
\end{align} 
which consists of the context information ${C_{\rm{p}}^\tau }$ and the query signal  ${{{\bf{y}}_{{\rm{query}}}}}$, which is  the received channel output. 

The goal of the ICL  denoiser is to estimate the transmitted symbol ${{{\bf{x}}_{{\rm{query}}}}}$ from the prompt 
${C_{\rm{p}}^\tau }$,  expressed as 
\begin{align}
	{{{\bf{\hat x}}}_{{\rm{query}}}} = {h_{\bm \omega} }\left( {C_{\rm{p}}^\tau ,{{\bf{y}}_{{\rm{query}}}}} \right), \label{ICL_estimatedsymbol}
\end{align}
where ${h_{\bm \omega} }\left(  \cdot  \right)$ denotes the mapping function parameterized by ${\bm \omega} $.

Based on \eqref{ICL_estimatedsymbol}, the training loss function is defined as the  mean squared error (MSE) averaged over all tasks and sequence symbols during pre-training, given by 
\begin{align}
{{\cal L}_{\rm ICL}}\left( {\bm \omega} \right) = \frac{1}{{(N + 1)M}}{{\mathbb E}_{\tau,{\bf x}_{{\rm p},n}} }\left[ {\sum\limits_{n = 1}^{N + 1} {\left\| {{{\bf{x}}_{{\rm{p}},n}} - {{{\bf{\hat x}}}_{{\rm{p}},n}}} \right\|_2^2} } \right], 
\end{align}
where ${{{\bf{\hat x}}}_{{\rm{p}},n}} = {h_{\bm \omega} }\left( {C_{\rm{p}}^\tau ,{{\bf{y}}_{{\rm{p}},n}}} \right), n \in \left[ {1, \ldots ,N} \right]$, ${{\bf{x}}_{{\rm{p}},N + 1}} = {{\bf{x}}_{{\rm{query}}}}$, and ${{{\bf{\hat x}}}_{{\rm{p}},N + 1}} = {{{\bf{\hat x}}}_{{\rm{query}}}}$. Note that the query signal  ${{\bf{x}}_{{\rm{query}}}}$ is available  during training but not during inference.
\subsection{Loss Function}
The ICL  denoiser, DeepJSCC encoder, and DeepJSCC decoder are jointly optimized to minimize the loss function given by
\begin{align}
{{\cal L}_{{{\rm tot}}}} = & \underbrace {\frac{1}{{CHW}}{{\mathbb E}_{\tau, \bf S} }\left[ {{{\left\| {{\bf{S}} - {\bf{\hat S}}} \right\|}^2}} \right]}_{{\kern 1pt} {\kern 1pt} {\rm{Image}}{\kern 1pt} {\kern 1pt} {\rm{reconstruction}}{\kern 1pt} {\kern 1pt} {\rm{loss}}} \notag\\
& +\lambda\underbrace {\frac{1}{{(N + 1)M}}{{\mathbb E}_{\tau,{\bf x}_{{\rm p},n}} }\left[ {\sum\limits_{n = 1}^{N + 1} {\left\| {{{\bf{x}}_{{\rm{p}},n}} - {{{\bf{\hat x}}}_{{\rm{p}},n}}} \right\|_2^2} } \right]}_{{\rm{ICL{\text -}based {\kern 1pt} {\kern 1pt}estimation{\kern 1pt} {\kern 1pt} loss}}} , \label{trainingloss}
\end{align}
where $\lambda $ denotes a non-negative weighting factor balancing the two loss components. The overall loss comprises two parts: the first term represents the end-to-end image reconstruction loss, averaged over channel realizations and input source images; the second term corresponds to the ICL-based estimation loss. Empirically, we observe that incorporating the second term significantly improves the convergence efficiency during training.

\section{ICL for DeepJSCC under Perfect IQ Balance}
In this section, we integrate the ICL  denoiser into DeepJSCC-based MIMO systems under the assumption of perfect IQ balance, i.e., ideal hardware without any distortions. Previous studies \cite{zecchin2024incontext,Zecchin2024cellfree,kunde2023transformers} typically considered exploiting pilot sequences and their outputs as context only for MIMO estimation via an ICL detector. In these works, the input signals belong to a finite constellation diagram, and the corresponding tasks are essentially symbol-detection problems. 
In contrast, our setting does not constrain the inputs to any finite alphabet, which makes the underlying inference task  more challenging.
We propose using the context information (pilot sequences and their outputs) not only in the   ICL denoiser  but also as input to both the DeepJSCC encoder and decoder. This allows the DeepJSCC encoder, decoder, and the ICL  denoiser to jointly optimize encoding, decoding, and estimation strategies for specific channel realizations.

\begin{figure}[!t]
	\centerline{\includegraphics[width=3.8in]{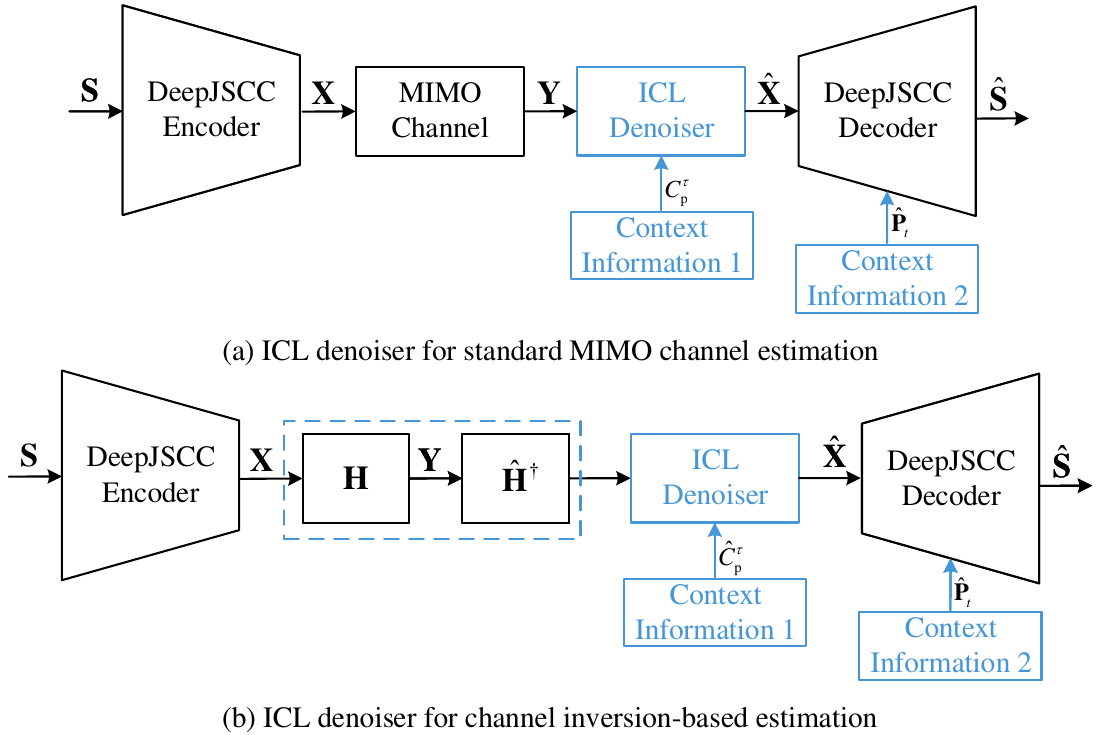}}
	\caption{Architecture of DeepJSCC-enabled image transmission over open-loop MIMO systems via ICL.} \label{Architecture_open_loop_icl}
%	\vspace{-0.3cm}
\end{figure}
\subsection{ICL  Denoiser for Open-Loop MIMO Systems}\label{ICL_Decoder_Open_Loop_Detection}
In this subsection, we investigate the integration of  ICL into open-loop MIMO systems, as illustrated in Fig.~\ref{Architecture_open_loop_icl}. We propose two  methods for the ICL  denoiser, differentiated by how channel estimation is performed. Below, we provide detailed descriptions of these two  methods.
\subsubsection{ICL  Denoiser for Standard  MIMO Channel Estimation}
In this scenario, the conventional MIMO channel estimation approach shown in Fig.~\ref{DeepJSCC}(a) is replaced by the ICL  denoiser. As shown in Fig.~\ref{Architecture_open_loop_icl}(a),
the ICL denoiser estimates the transmitted symbols $\bf X$ based on the context ${C_{\rm{p}}^\tau }$, termed as  \textit{context information 1}. According to \eqref{ICL_estimatedsymbol}, the estimate of $\bf X$, denoted by ${{\bf{\hat X}}}$, can be expressed as 
\begin{align}
{\bf{\hat X}} = {h_{\bm \omega} }\left( {C_{\rm{p}}^\tau ,{\bf{Y}}} \right).
\end{align}
We remind that  channel $\bf H$ is assumed to remain constant throughout the transmission of one image, corresponding to $L$ channel uses. 

In addition, we also utilize the pilot sequences and their outputs $C_{\rm{p}}^\tau $ to generate context information for the DeepJSCC decoder. We term this context information as \textit{context information 2}. Two types of \textit{context information 2} is considered. One is channel heatmap  and the other is ICAR, which are described as below.

\textbf{Channel Heatmap Construction:} An intuitive approach is to utilize the estimated CSI and noise variance as contextual information for the DeepJSCC decoder. This enables the decoder to infer the current channel conditions, identify more reliable sub-channels and transmit more critical source feature through the more favorable subchannels. To begin with, we estimate the channel matrix 
$\bf H$ from the pilot observations $C_{\rm{p}}^\tau $. Specifically, we organize the received pilot sequences into a matrix as follows:

\begin{align}
	{{\bf{Y}}_{\rm{p}}} = {\bf{H}}{{\bf{X}}_{\rm{p}}} + {{\bf{W}}_{\rm{p}}}, \label{pilotequation}
\end{align}
where ${{\bf{Y}}_{\rm{p}}} = \left[ {{{\bf{y}}_{{\rm{p}},1}}, \ldots ,{{\bf{y}}_{{\rm{p}},N}}} \right]$, ${{\bf{X}}_{\rm{p}}} = \left[ {{{\bf{x}}_{{\rm{p}},1}}, \ldots ,{{\bf{x}}_{{\rm{p}},N}}} \right]$, and ${{\bf{W}}_{\rm{p}}} = \left[ {{{\bf{w}}_{{\rm{p}},1}}, \ldots ,{{\bf{w}}_{{\rm{p}},N}}} \right]$.
We consider the LS channel estimation method. Although MMSE generally provides better estimation accuracy at lower SNR, it requires statistical knowledge of both the channel and noise. In contrast, LS estimation does not require statistical channel or noise information, making it simpler and more practical. Based on \eqref{pilotequation}, we adopt the LS method for estimating the channel as
\begin{align}
	\widehat {\bf{H}} = {{\bf{Y}}_{\rm{p}}}{({\bf{X}}_{\rm{p}}^H{{\bf{X}}_{\rm{p}}})^{ - 1}}{\bf{X}}_{\rm{p}}^H.\label{Estimation_H}
\end{align}
After obtaining the estimated channel $\widehat {\bf{H}}$, 
 the channel heatmap, denoted by  ${\bf{\hat P}}_{\rm CHM}\in {{\mathbb R}^{M \times M}}$, is defined as 
\begin{align}
{\bf{\hat P}}_{\rm CHM} = {\sigma ^2}\left( {{{\widehat {\bf{H}}}_{{\rm{ZF}}}} \odot \widehat {\bf{H}}_{_{{\rm{ZF}}}}^*} \right), \label{open_loop_channelheatmap}
\end{align}
where ${\widehat {\bf{H}}_{{\rm{ZF}}}} = {\left( {{{\bf{H}}^H}{\bf{H}}} \right)^{ - 1}}{{\bf{H}}^H}$. Note that the noise variance is assumed to be known, as it can be estimated during an offline phase. From \eqref{open_loop_channelheatmap}, the channel heatmap encodes the effective noise variance and degradation experienced across different spatial and temporal transmission paths, allowing the decoder to adaptively adjust its attention across the received signals.

\noindent\textbf{Implicit Channel-Aware Representation:} Although a simple channel heatmap can convey the underlying channel conditions, it relies on an accurate estimation of the CSI. However, in scenarios where the channel undergoes deep fading or exhibits strong non-linear characteristics, conventional LS-based CSI estimation may fail. To address this issue,  we propose using ICAR which employs neural networks to learn implicit representations of the channel without explicitly constructing a channel heatmap. 
In this approach, the pilot sequences and their corresponding  outputs $C_{\rm{p}}^\tau $  are utilized to extract latent CSI features via neural networks.  Specifically, the channel heatmap is replaced with two fully connected (FC) layers, each followed by a ReLU activation function. The dimensions of the first and second FC layers are $2N\times 4N$ and  $4N\times M$, respectively. Since both the pilot sequences and their outputs are complex-valued, we first convert them into real-valued representations. For each $n \in \left[ {1, \ldots ,N} \right]$, we define 
\begin{align}
&{{{\bf{\bar x}}}_{{\rm{p}},n}} = {\left[ {{\mathop{\rm Re}\nolimits} \left\{ {{\bf{x}}_{{\rm{p}},n}^T} \right\},{\mathop{\rm Im}\nolimits} \left\{ {{\bf{x}}_{{\rm{p}},n}^T} \right\}} \right]^T}, \notag\\
&{{{\bf{\bar y}}}_{{\rm{p}},n}} = {\left[ {{\mathop{\rm Re}\nolimits} \left\{ {{\bf{y}}_{{\rm{p}},n}^T} \right\},{\mathop{\rm Im}\nolimits} \left\{ {{\bf{y}}_{{\rm{p}},n}^T} \right\}} \right]^T}.
\end{align}
These real-valued vectors are then concatenated to form a matrix of size $2M\times 2N$,   which serves as the input to the FC networks. The output of the FC networks is denoted by ${{{\bf{\hat P}}}_{{\rm{ICAR}}}} \in {{\mathbb C}^{2M \times M}}$.

Finally, the DeepJSCC decoder takes both ${{\bf{\hat P}}_{ t}}, t \in \left\{ {{\rm{CHM,ICAR}}} \right\},$ and ${{\bf{\hat X}}}$ as input. The reconstructed image is then given by 
\begin{align}
{\bf{\hat S}} = {g_{\bm \phi} }\left( {{\bf{\hat X}},{\bf{\hat P}}_t} \right).
\end{align}
 A simple approach for combining these inputs is directly concatenating them, after performing complex-to-real conversion and reshaping operations to match the input dimensions required by the DeepJSCC decoder. To be specific, we first reshape the complex matrix ${{\bf{\hat X}}}$ into a real matrix  
${\bf{\bar X}} \in {{\mathbb R}^{p^2 \times \frac{{2ML}}{{{p^2}}}}}$, matching the output dimensions of the DeepJSCC encoder. For  channel  heatmap ${\bf{\hat P}}_{\rm CHM} $, it
undergoes reshaping and repetition operations as
\begin{align}
{\bf{\hat P}}_{\rm CHM} \in {{\mathbb R}^{M \times M}}&\xrightarrow{\text{reshape}}{\bf{\tilde P}}_{\rm CHM} \in {{\mathbb R}^{{M^2} \times 1}}\xrightarrow{\text{repeat}} \notag\\
&{\bf{\bar P}}_{\rm CHM} \in {{\mathbb R}^{{p^2} \times {M^2}}}. \label{reshaped_channelheatmap}
\end{align}
Similary, for ICAR, we transform ${\bf{\hat P}}_{\rm ICAR} $ as 
\begin{align}
{\bf{\hat P}}_{\rm ICAR} \in {{\mathbb R}^{2M \times M}}&\xrightarrow{\text{reshape}}{\bf{\tilde P}}_{\rm ICAR} \in {{\mathbb R}^{{2M^2} \times 1}}\xrightarrow{\text{repeat}} \notag\\
&{\bf{\bar P}}_{\rm ICAR} \in {{\mathbb R}^{{p^2} \times {2M^2}}}. \label{reshaped_ICAR}
\end{align}

Then, we concatenate ${{\bf{\bar X}}}$ and ${{\bf{\bar P}}_{t}}$ as 
\begin{align}
{\bf{X}}_{{\rm{decoder}}}^{{\rm{in}}} = \left\{ \begin{array}{l}
{\rm{Concat}}\left( {{\bf{\bar X}},{{{\bf{\bar P}}}_{{\rm{CHM}}}}} \right) \in {{\mathbb R}^{{p^2} \times \left( {\frac{{2ML}}{{{p^2}}} + {M^2}} \right)}},\\
{\rm{Concat}}\left( {{\bf{\bar X}},{{{\bf{\bar P}}}_{{\rm{ICAR}}}}} \right) \in {{\mathbb R}^{{p^2} \times \left( {\frac{{2ML}}{{{p^2}}} + 2{M^2}} \right)}}.
\end{array} \right. \label{TX:combinedtwosignals}
\end{align}
where ${\rm{Concat}}\left(  \cdot  \right)$ denotes the concatenation operation. This combined input  ${\bf{X}}_{{\rm{decoder}}}^{{\rm{in}}}$ is then fed into the DeepJSCC decoder.

\subsubsection{ICL  Denoiser for Channel Inversion-based Estimation} In this scenario, we propose a channel inversion-based architecture for open-loop systems as depicted in Fig.~\ref{Architecture_open_loop_icl}(b).
This architecture is similar to the previously described one, except that channel inversion is applied to simplify MIMO estimation, enabling the ICL denoiser to more effectively estimate the transmitted symbols.  Specifically, the received signal $\bf Y$ first passes through a channel inversion module, represented by ${{\bf{\hat H}}^\dag }$, computed according to  \eqref{Estimation_H}. Under this configuration, the ICL  denoiser learns the mapping function  ${\bf{H}}{{\bf{\hat H}}^\dag }$, which ideally simplifies to the identity matrix $\bf I$ if $\bf H$  is full-rank and perfectly estimated, i.e.,
\begin{align}
	{\bf{H}}{{\bf{\hat H}}^\dag } = {\bf{I}}.
\end{align}
This structure effectively prevents signal overlap from different antennas. Under this scenario, the estimation by the ICL  denoiser can be represented as: 
\begin{align}
{\bf{\hat X}} = {h_{\bm \omega} }\left( { \hat C_{\rm{p}}^\tau ,{{\bf H}^{\dag}}{\bf Y}} \right),
\end{align}
where $\hat C_{\rm{p}}^\tau  = \left\{ {\left( {{{{\bf{\hat y}}}_{{\rm{p}},1}},{{\bf{x}}_{{\rm{p}},1}}} \right), \ldots ,\left( {{{{\bf{\hat y}}}_{{\rm{p}},N}},{{\bf{x}}_{{\rm{p}},N}}} \right)} \right\}$, and ${{{{\bf{\hat y}}}_{{\rm{p}},n}}}$ is the $n$-th column vector of ${{{\bf{\hat Y}}}_{\rm{p}}} = {{\bf{H}}^\dag }{{\bf{Y}}_{\rm{p}}}$.

Finally, we combine ${{\bf{\hat P}}_t}$ and ${{\bf{\hat X}}}$ similarly to \eqref{TX:combinedtwosignals}, and  feed them into the DeepJSCC decoder.
\begin{figure}[!t]
	\centerline{\includegraphics[width=3.8in]{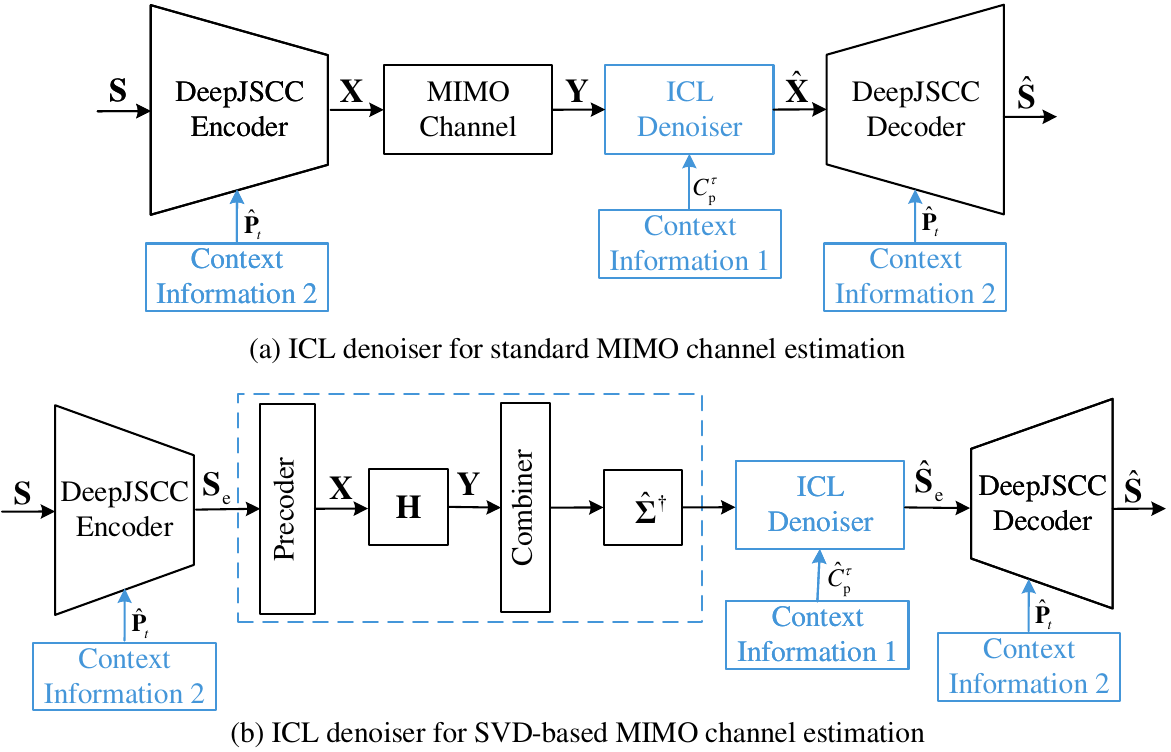}}
	\caption{Architecture of DeepJSCC-enabled image transmission over closed-loop MIMO systems via ICL.} \label{Architecture_closed_loop_icl}
%	\vspace{-0.3cm}
\end{figure}
\subsection{ICL  Denoiser for Closed-Loop MIMO  Systems}\label{ICL_Decoder_closed_Loop_Detection}
In this subsection, we explore DeepJSCC-enabled image transmission in closed-loop MIMO systems using the ICL  denoiser, as depicted in  Fig.~\ref{Architecture_closed_loop_icl}. 
We consider two scenarios: one employing a standard MIMO estimation approach and the other employing an SVD-based MIMO estimation.
\subsubsection{ICL  Denoiser for Standard  MIMO Channel Estimation} As illustrated in Fig.~\ref{Architecture_closed_loop_icl}(a),  \textit{context information 2}, i.e., channel heatmap or ICAR, is available to both the DeepJSCC encoder and decoder. 
The overall architecture of Fig.~\ref{Architecture_closed_loop_icl}(a) is similar to that of Fig.~\ref{Architecture_open_loop_icl}, with the main difference being the provision of additional context information to the DeepJSCC encoder.
The process of concatenating the decoded symbols ${\bf{\hat X}}$ with the channel heatmap or ICAR follows the same approach detailed in
 Section \ref{ICL_Decoder_Open_Loop_Detection}.

Next, we describe how the input image $\bf S$ and channel heatmap ${\bf{\hat P}}_t$ are concatenated.  Before feeding image $\bf S$ into the DeepJSCC encoder, it is divided into a grid of $p\times p$ patches, each reshaped into a vector of dimension  ${\frac{{CHW}}{{{p^2}}}}$.  Consequently, the image $\bf S$ is transformed into a  matrix, denoted by  ${{\bf{S}}_{\rm{p}}} \in {{\mathbb R}^{{p^2} \times \frac{{CHW}}{{{p^2}}}}}$, where ${{p^2}}$ represents the total number of patches and 
 ${\frac{{CHW}}{{{p^2}}}}$  denotes the dimensionality of each patch.
 
Based on ${{\bf{S}}_{\rm{p}}}$, \eqref{reshaped_channelheatmap}, and \eqref{reshaped_ICAR}, the input to the DeepJSCC encoder is formulated as
\begin{align}
{\bf{X}}_{{\rm{encoder}}}^{{\rm{in}}} = \left\{ \begin{array}{l}
{\rm{Concat}}\left( {{{\bf{S}}_{\rm{p}}},{{{\bf{\bar P}}}_{{\rm{CHM}}}}} \right) \in {{\mathbb R}^{{p^2} \times \left( {\frac{{CHW}}{{{p^2}}} + {M^2}} \right)}},\\
{\rm{Concat}}\left( {{{\bf{S}}_{\rm{p}}},{{{\bf{\bar P}}}_{{\rm{ICAR}}}}} \right) \in {{\mathbb R}^{{p^2} \times \left( {\frac{{CHW}}{{{p^2}}} + 2{M^2}} \right)}}.
\end{array} \right. 
\end{align}

\subsubsection{ICL  Denoiser for SVD-based  MIMO Channel Estimation} 
In this scenario, we consider SVD-based transmission, where the ICL denoiser learns the mapping function ${{{\bf{\hat \Sigma }}}^\dag }{{{\bf{\hat U}}}^H}{\bf{H\hat V}}$, with the estimated channel represented as ${\bf{\hat H}} = {\bf{\hat U\hat \Sigma }}{{{\bf{\hat V}}}^H}$. In the ideal case, where CSI is perfectly estimated, i.e., ${\bf{\hat H}} = {\bf{H}}$, this function simplifies to ${{{\bf{\hat \Sigma }}}^\dag }{{{\bf{\hat U}}}^H}{\bf{H\hat V}} = {\bf{I}}$.
This structure effectively parallelizes data transmission and prevents signal overlap from different antennas. Under this scenario, the estimation by the ICL  denoiser can be represented as: 
\begin{align}
{\bf{\hat X}} = {h_{\bm \omega} }\left( {\hat C_{\rm{p}}^\tau ,{{{\bf{\hat \Sigma }}}^\dag }{{{\bf{\hat U}}}^H}{\bf{Y}}} \right),
\end{align}
where $\hat C_{\rm{p}}^\tau  = \left\{ {\left( {{{{\bf{\hat y}}}_{{\rm{p}},1}},{{\bf{x}}_{{\rm{p}},1}}} \right), \ldots ,\left( {{{{\bf{\hat y}}}_{{\rm{p}},N}},{{\bf{x}}_{{\rm{p}},N}}} \right)} \right\}$, and ${{{{\bf{\hat y}}}_{{\rm{p}},n}}}$ is the $n$-th column vector of ${{{\bf{\hat Y}}}_{\rm{p}}} = {{{\bf{\hat \Sigma }}}^\dag }{{{\bf{\hat U}}}^H}{{\bf{Y}}_{\rm{p}}}$.
Compared to the scenario without SVD-based MIMO estimation, the mapping function in the SVD-based approach is simpler to learn. 

In addition,
since SVD is performed, the effective noise level after MIMO equalization in closed-loop systems differs from that in open-loop systems, indicating that the channel heatmap computed by \eqref{open_loop_channelheatmap}  is no longer applicable. According to \eqref{closed_loop_detection}, the effective noise after MIMO equalization can be expressed as ${{\bf{\Sigma }}^\dag }{{\bf{U}}^H}{\bf{W}}$.  Thus, the channel heatmap for the closed-loop system is redefined as
\begin{align}
{\bf{\hat P}}_{\rm CHM} = {\sigma ^2}\left( {\left( {{{{\bf{\hat \Sigma }}}^\dag }{{{\bf{\hat U}}}^H}} \right) \odot {{\left( {{{{\bf{\hat \Sigma }}}^\dag }{{{\bf{\hat U}}}^H}} \right)}^*}} \right),
\end{align}
where ${{\bf{\hat \Sigma }}}$ and ${{\bf{\hat U}}}$ are obtained from the SVD of the estimated channel ${\bf{\hat H}}$.
  Then, we incorporate this channel heatmap as supplementary context information for both the DeepJSCC encoder and decoder.  The concatenation of the input image and the channel heatmap follows the same procedure as previously described and has been omitted for brevity. Note that for ICAR, it follows the same procedures for the standard MIMO channel estimation for closed-loop systems.

\textbf{\textit{Remark:}} A trade-off exists between feedback overhead and system performance when utilizing channel heatmap and ICAR. While transmitting pilot sequences along with their corresponding outputs incurs substantially higher overhead compared to the channel heatmap, the ICAR framework offers a superior capability to implicitly capture channel characteristics, thereby enabling enhanced system performance.

\begin{figure}[!t]
	\centerline{\includegraphics[width=3.6in]{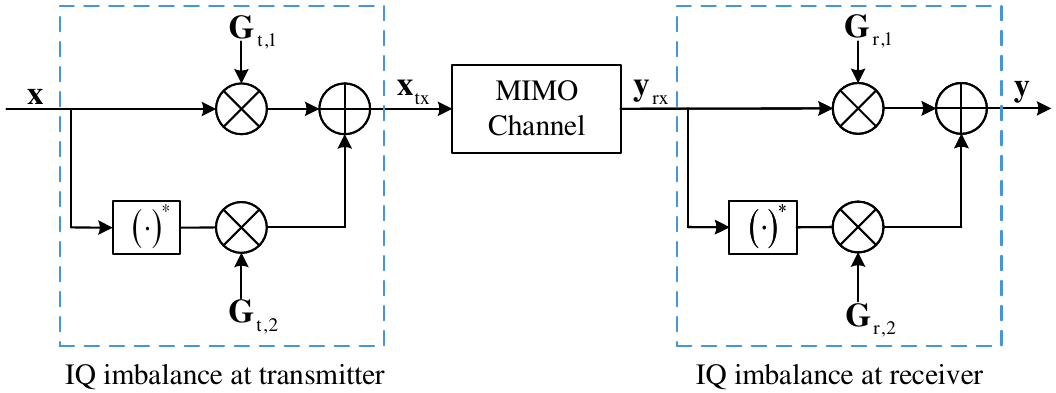}}
	\caption{Architecture of IQ imbalance at both the transmitter and receiver.} \label{IQimbalance:Architecture}
%	\vspace{-0.3cm}
\end{figure}
  
\section{ICL for DeepJSCC under IQ Imbalance}
\subsection{IQ Imbalance}
We consider a general scenario where IQ imbalance occurs at both the transmitter and receiver due to the hardware impairment, such as mixers, amplifiers, filters, etc, causing amplitude and phase mismatches \cite{soleymani2022improper,javed2019multiple,Soleymani2020improperx}. The structure of IQ imbalance is illustrated in Fig.~\ref{IQimbalance:Architecture}. Let $\bf x$ represent the transmitted information signal. After passing through imperfect hardware, the distorted transmitted signal ${{\bf{x}}_{{\rm{tx}}}}$ can be expressed as 
\begin{align}
{{\bf{x}}_{{\rm{tx}}}} = {{\bf{G}}_{{\rm{t}},1}}{\bf{x}} + {{\bf{G}}_{{\rm{t}},2}}{{\bf{x}}^*}, \label{IQ_at_transmitter}
\end{align}
where the matrices ${{\bf{G}}_{{\rm{t}},1}}$ and ${{\bf{G}}_{{\rm{t}},2}}$ capture amplitude and phase mismatches, respectively, and are defined as 
\begin{align}
{{\bf{G}}_{{\rm{t}},1}} = \frac{{{\bf{I}} + {{\bf{A}}_t}{e^{j{{\bm{\Theta }}_{\rm{t}}}}}}}{2},{{\bf{G}}_{{\rm{t}},2}} = \frac{{{\bf{I}} - {{\bf{A}}_t}{e^{ - j{{\bm{\Theta }}_{\rm{t}}}}}}}{2},
\end{align}
with ${\bf{I}}$ being the identity matrix, and diagonal matrices ${{{\bf{A}}_t}}$ and ${{{\bf{\Theta }}_{\rm{t}}}}$ modeling amplitude and phase errors for each transmitter branch, respectively. Note that, 
in an ideal IQ scenario (perfect IQ balance), we have ${{\bf{A}}_t} = {\bf{I}}$ and ${{\bf{\Theta }}_{\rm{t}}} = {\bf{0}}$, resulting in ${{\bf{x}}_{{\rm{tx}}}} = {\bf{x}}$.

Similarly, at the receiver side, the received signal ${\bf{y}}_{\rm rx}$ after passing through imperfect hardware is modeled as 
\begin{align}
{{\bf{y}}} = {{\bf{G}}_{{\rm{r}},1}}{\bf{y}}_{\rm rx} + {{\bf{G}}_{{\rm{r}},2}}{{\bf{y}}_{\rm rx}^*}, \label{IQ_at_receiver}
\end{align}
where ${{\bf{G}}_{{\rm{r}},1}} = \frac{{{\bf{I}} + {{\bf{A}}_r}{e^{j{{\bm{\Theta }}_{\rm{r}}}}}}}{2}$ and ${{\bf{G}}_{{\rm{r}},2}} = \frac{{{\bf{I}} - {{\bf{A}}_r}{e^{ - j{{\bm{\Theta }}_{\rm{r}}}}}}}{2}$, which are similar to the IQ imbalance at the transmitter.  In the ideal case at the receiver, ${{\bf{A}}_r} = {\bf{I}}$ and ${{\bf{\Theta }}_{\rm{r}}} = {\bf{0}}$.

Based on \eqref{IQ_at_transmitter}, \eqref{IQ_at_receiver}, and ${{\bf{y}}_{{\rm{rx}}}} = {\bf{H}}{{\bf{x}}_{{\rm{tx}}}} + {\bf{w}}$, it is evident that the final received signal $\bf y$ becomes a nonlinear function of the originally transmitted symbol $\bf x$, due to the distortion introduced by IQ imbalance. 

%\begin{figure}[!t]
%	\centerline{\includegraphics[width=3.2in]{ICL_decoder_open_loop_architecture.eps}}
%	\caption{ICL-based estimation for open-loop MIMO systems with IQ imbalance.} \label{ICL_decoder_open_loop_architecture}
%	\vspace{-0.3cm}
%\end{figure}
\subsection{Open-Loop MIMO  Systems with IQ Imbalance}\label{ICL_Decoder_Open_Loop_Detection_IQI}
In this subsection, we examine open-loop MIMO systems with IQ imbalance at both the transmitter and receiver sides. ICL  denoiser is applied to handle the nonlinear MIMO estimation problem. Let  blocks ${{\bf{G}}_{\rm{t}}}$ and ${{\bf{G}}_{\rm{r}}}$ represent the transmitter and receiver IQ imbalance modules depicted in Fig.~\ref{IQimbalance:Architecture}. 
 Compared to the ICL-based estimation under the perfect IQ balance scenario described previously in Fig.~\ref{Architecture_open_loop_icl}(a), here the ICL denoiser learns a nonlinear mapping function from the transmitter through the channel and receiver impairments, i.e., ${{\bf{G}}_{\rm{t}}} \to {\bf{H}} \to {{\bf{G}}_{\rm{r}}}$. 
  In addition, compared to the ICL denoiser for channel inversion-based estimation under the perfect IQ balance scenario  in  Fig.~\ref{Architecture_open_loop_icl}(b),  the ICL denoiser learns a nonlinear mapping  from the transmitter through the channel and receiver impairments, i.e., ${{\bf{G}}_{\rm{t}}} \to {\bf{H}} \to {{\bf{G}}_{\rm{r}}}\to {{{\bf{\tilde H}}}^\dag }$, where ${{\bf{\tilde H}}}$ is obtained by applying the LS method to approximate the nonlinear end-to-end mapping characterized by ${{\bf{G}}_{\rm{t}}} \to {\bf{H}} \to {{\bf{G}}_{\rm{r}}}$.
 It is important to note that the final channel inversion module ${{{\bf{\tilde H}}}^\dag }$  represents the inversion of the non-linear MIMO channel, and therefore cannot fully capture the true nonlinear mapping introduced by IQ imbalance. Consequently, this channel inversion-based approach may lead to degraded performance, a point further investigated in our simulation results. Moreover, the channel heatmap obtained from ${{\bf{\tilde H}}}$ may significantly deviate from true channel conditions, further degrading its performance.  In contrast, the ICAR as contextual information for DeepJSCC decoder could  learn implicit representations of the distorted channel conditions directly from pilot responses using neural networks, avoiding the need for an explicitly constructed heatmap. These interesting findings have been revealed through our simulations.
 
%\begin{figure}[!t]
%	\centerline{\includegraphics[width=3.8in]{ICL_decoder_closed_loop_architecture.eps}}
%	\caption{ICL-based estimation for closed-loop MIMO systems with IQ imbalance.} \label{ICL_decoder_closed_loop_architecture}
%	\vspace{-0.3cm}
%\end{figure}
\subsection{Closed-Loop MIMO  Systems with IQ Imbalance}
In this subsection, we investigate the ICL-based estimation for SVD-based closed-loop MIMO systems affected by IQ imbalance.  Similarly to  Fig.~\eqref{Architecture_closed_loop_icl},  we  consider two architectures that incorporate IQ imbalance into the design. For the standard MIMO channel estimation scenario, the system follows the structure ${\bf{X}} \to {{\bf{G}}_{\rm{t}}} \to {\bf{H}} \to {{\bf{G}}_{\rm{r}}} \to {\bf{\hat X}}$. For the SVD-based MIMO channel estimation scenario, the input signal is first processed by the estimated precoding matrix ${{\bf{\tilde V}}}$, followed by IQ distortion  components ${\bf G}_{\rm t}$ and ${\bf G}_{\rm r}$, and channel $\bf H$, where ${\bf{\tilde H}} = {\bf{\tilde U\tilde \Sigma }}{{{\bf{\tilde V}}}^H}$ is derived by performing SVD. The received signal is subsequently processed via ${{{\bf{\tilde U}}}^H}$ and ${{\bf{\tilde \Sigma }}}^{\dag}$.  However, due to the presence of IQ imbalance, the overall transformation  ${\bf{\tilde V}} \to {{\bf{G}}_{\rm{t}}} \to {\bf{H}} \to {{\bf{G}}_{\rm{r}}} \to {{{\bf{\tilde U}}}^H} \to {{{\bf{\tilde \Sigma }}}^\dag }$ becomes highly nonlinear. Note that for both designs, the ICL denoiser is trained to learn the overall nonlinear mapping from transmitter to receiver, capturing the joint impact of the physical channel and IQ impairments.

 In addition, we incorporate both the channel heatmap or the proposed ICAR module for both the DeepJSCC encoder and decoder. Experimental results  demonstrate that the ICAR-enhanced ICL denoiser offers superior robustness and adaptability, particularly in nonlinear and imperfect CSI regimes, making it a viable alternative to traditional closed-loop methods.

\section{Numerical Results}
In this section, we present numerical experiments to validate the effectiveness of the proposed DeepJSCC-based  enhanced by the transformer-based ICL  and additional contextual information in MIMO systems.  In the literature, various architectures for DeepJSCC encoders and decoders have been studied previously,  including  CNN-based and transformer-based frameworks. Here, we adopt a ViT-based architecture, following the design in \cite{wu2024deepmimo}. This choice is motivated by the superior ability of ViT to capture long-range dependencies across the entire input, whereas CNN-based architectures only rely on local receptive fields.\footnote{Note that this paper primarily focuses on how to design the ICL  and contextual information to enhance image reconstruction performance, rather than developing the DeepJSCC encoder/decoder architectures.}

The input source images are selected from the CIFAR-10 dataset \cite{cukierski2013cifar10}, which consists of 60000 color images of size $3\times32\times32$ (color channels, height, width) in 10 classes, with 6000 images per class.  The training and evaluation datasets from  CIFAR-10 consist of distinct images, containing  50000 and 10000 images, respectively. All results are averaged over independent channel realizations for each image in the evaluation dataset.

The model is implemented in PyTorch and optimized using the Adam optimizer. The learning rate is set to $10^{-4}$ with a batch size of $64$.  Unless stated otherwise, the number of transmit antennas is set to $M=2$, corresponding to a $2\times2$ MIMO system. The number of channel uses is set to $L=256$, resulting in a bandwidth ratio of  $\frac{L}{{CHW}} = \frac{1}{{12}}$. In addition, we set $\lambda=0.01 $ and $\sigma^2=1$. To balance computational complexity and system performance, each input image is partitioned into an  $8\times8$ grid of patches.

To evaluate the performance of our proposed scheme, we adopt the classic peak signal-to-noise ratio (PSNR) metric for image reconstruction. The PSNR expresses the ratio between the maximum possible power of a signal and the power of corrupting noise that affects the fidelity of its representation, which is  given by 
\begin{align}
\text{PSNR} \triangleq 10 \log_{10} \left( \frac{{\rm MAX}_{\rm I}^2}{\text{MSE}({\bf{S}}, {{\bf{\hat S}}})} \right)~  (\rm{dB}), \label{PSNR}
\end{align}
where ${\rm MAX}_{\rm I}$  is the maximum possible pixel value of the image, which is 255
 as we use 24-bit depth RGB images, and ${\rm{MSE}}\left( {{\bf{S}},{\bf{\hat S}}} \right)  \triangleq  \frac{1}{{CHW}}\left\| {{\bf{S}} - {\bf{\hat S}}} \right\|^2$ stands for the MSE between ${\bf{S}}$ and  ${{\bf{\hat S}}}$. From \eqref{PSNR},  we observe that a higher PSNR indicates better image reconstruction quality.

The channel quality is characterized by the transmit SNR, defined as 
\begin{align}
	{\rm{SNR}} \triangleq  10{\log _{10}}\left( {\frac{P}{{{\sigma ^2}}}} \right)~  \left( {{\rm{dB}}} \right).
\end{align}
In addition, we use a decoder-only transformer architecture \cite{vaswani2017attention} from the GPT-2 family \cite{radford2018improving} as the backbone of the ICL  denoiser. The model consists of 2 attention layers, 4 attention heads, and a 64-dimensional embedding space.
Transmitted pilots are independently sampled from a Gaussian distribution, i.e.,  ${{\bf{x}}_{{\rm{p}},n}} \sim {\cal CN}\left( {{\bf{0}},{\bf{I}}} \right),\forall n$, and normalized to satisfy the power constraint before transmission.

\subsection{ICL  Denoiser for Transmit Symbol Estimation over MIMO Systems}
\begin{figure}[!t]
	\centerline{\includegraphics[width=3.5in]{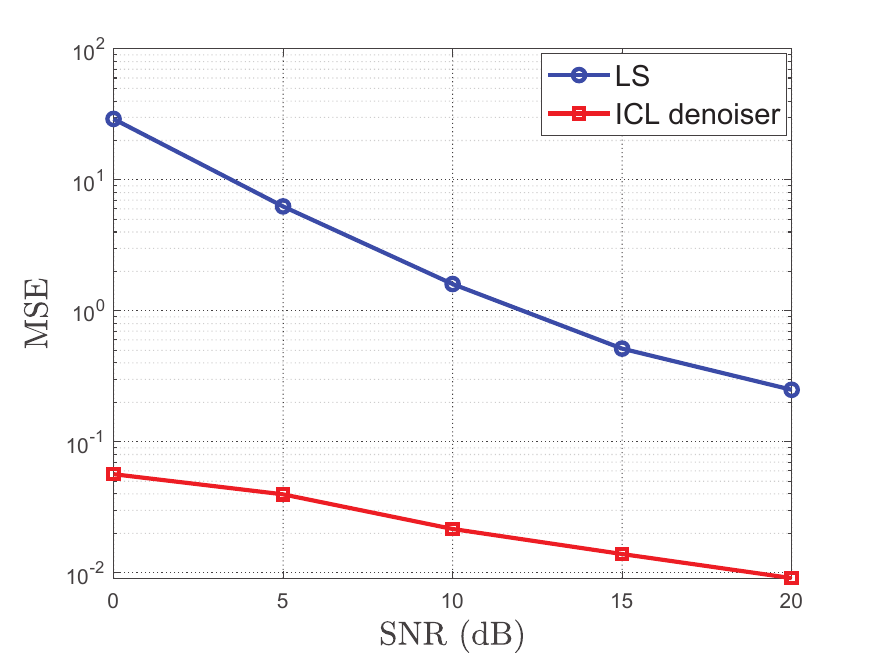}}
	\caption{MSE versus SNR  under the perfect IQ case with  pilot sequence length $N=11$.} \label{MSEvsSNR}
%	\vspace{-0.3cm}
\end{figure}
\begin{figure}[!t]
	\centerline{\includegraphics[width=3.5in]{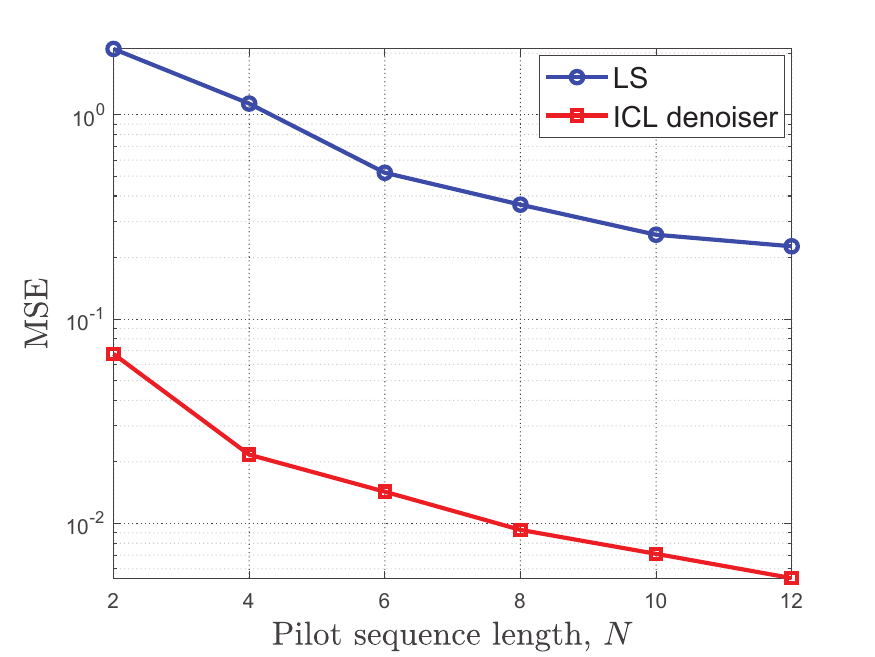}}
	\caption{MSE versus pilot sequence length  under the perfect IQ case with ${\rm SNR}=20~$dB.} \label{MSEvspilotlength}
%	\vspace{-0.3cm}
\end{figure}
\subsubsection{ICL  Denoiser for Inverse Linear Problem} We first evaluate a scenario without hardware impairments, where symbol estimation is an \textit{inverse linear problem}.
In Fig.~\ref{MSEvsSNR}, we compare the MSE performance of different decoding schemes as a function of  SNR under perfect IQ conditions, using a pilot sequence length of  $N=11$. 
The ``ICL denoiser" refers to the ICL-based decoding scheme described in Section \ref{ICL_architecture_model}, while the ``LS'' scheme denotes the conventional LS method used for symbol estimation. It is observed that the MSE achieved by the ICL  denoiser monotonically decreases with increasing SNR, confirming its ability to effectively learn the underlying linear function. Furthermore, the ICL  denoiser consistently achieves a lower MSE than the LS scheme. This performance gain can be attributed to two main reasons. First, the LS scheme estimates CSI based on pilot signals and then applies LS to recover the transmitted symbols from the noisy channel output and previously estimated CSI. This two-step estimation process introduces compounded errors, ultimately degrading accuracy. In contrast, the ICL denoiser does not require explicit CSI estimation. Instead, it directly recovers the channel input removing the noise \cite{kunde2023transformers}.

In Fig.~\ref{MSEvspilotlength},  the relationship between MSE and pilot sequence length $N$ is analyzed. It is observed that the MSE for both the ICL and LS schemes decrease as $N$ increases. This trend is expected, as a longer pilot sequence provides more information about the channel, improving the ICL denoiser’s ability to infer channel characteristics and thus enhancing symbol estimation accuracy. Notably, across all values of $N$, the ICL  denoiser consistently outperforms the LS scheme, further demonstrating its superior estimation capability.

\begin{figure}[!t]
	\centerline{\includegraphics[width=3.5in]{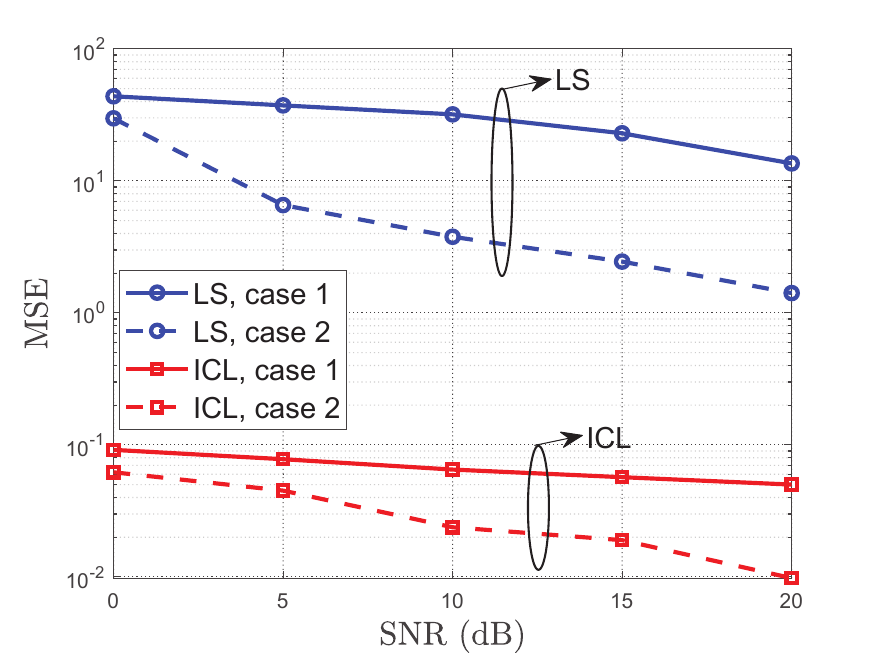}}
	\caption{MSE versus SNR  under the IQ imbalance case with  pilot sequence length $N=11$.} \label{IQI_MSEVsSNR}
%	\vspace{-0.3cm}
\end{figure}
\begin{figure}[!t]
	\centerline{\includegraphics[width=3.5in]{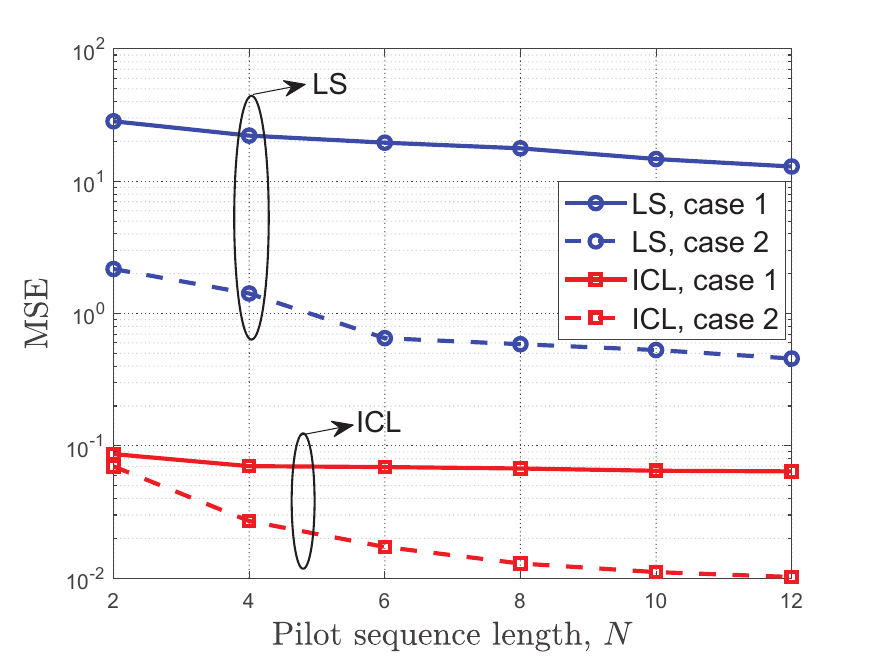}}
	\caption{MSE versus pilot sequence length  under the  IQ imbalance case with ${\rm SNR}=20~$dB.} \label{IQI_MSEVSpilotlength}
%	\vspace{-0.3cm}
\end{figure}
\subsubsection{ICL Denoiser for Inverse Non-Linear Problem}Next, we consider a more practical scenario where both the transmitter and receiver suffer from hardware imperfections, resulting in an \textit{inverse non-linear problem}. This introduces a more challenging symbol estimation task. We examine two hardware impairment scenarios, classified based on the severity of IQ imbalance. For Case 1, we assume that the hardware is significantly distorted. Specifically, each diagonal element of matrices  ${{\mathbf{A}}_t}$ and ${{\mathbf{A}}_r}$ follows a uniform distribution $\left[ {0,1} \right]$, and each diagonal element of 
${{\bm {\Phi}}_t} $ and ${{\bm \Phi}_r} $ follows a uniform distribution $\left[ {0,2\pi} \right]$. For Case 2, representing less severe hardware distortions, each diagonal element of ${{\mathbf{A}}_t}$ and ${{\mathbf{A}}_r}$ follows a  uniform distribution $\left[ {0.8,1} \right]$, and each diagonal element of 
${{\bm {\Phi}}_t} $ and ${{\bm \Phi}_r} $ follows a uniform distribution $\left[ {0,\frac{\pi }{{12}}} \right]$.

In Fig.~\ref{IQI_MSEVsSNR}, we investigate the MSE performance versus SNR under IQ imbalance, with a pilot sequence length of $N=11$. The results show that for Case 2 (i.e., less hardware distortion), the MSE achieved by the ICL denoiser decreases rapidly as SNR increases, indicating that the denoiser performs well even under non-linear channel conditions. For Case 1 (i.e., significant hardware distortion), although the MSE also decreases with increasing SNR, the reduction is more gradual, even at high SNR levels. Notably, across all SNR values, the ICL  denoiser consistently outperforms the LS scheme. For instance, at ${\rm SNR}=20~$dB for Case 1, the MSE achieved by the LS scheme is approximately $13.58$, while the ICL  denoiser achieves an MSE of only about $0.05$. This highlights that the conventional LS scheme fails under severe hardware impairments, whereas the ICL  denoiser maintains robust performance.

In Fig.~\ref{IQI_MSEVSpilotlength}, we further analyze the MSE versus pilot sequence length $N$ under the IQ imbalance condition, with the SNR fixed at 
${\rm SNR}=20~$dB. It is observed that the ICL  denoiser consistently outperforms the LS scheme across all values of $N$. As expected, both schemes achieve lower MSEs for Case 2 compared to Case 1.
However, for Case 1, the LS scheme fails to achieve satisfactory performance even as $N$ becomes large, whereas the ICL denoiser maintains effective estimation even with a small $N$. Additionally, we find that increasing 
$N$ significantly improves the MSE for Case 2, but has a negligible impact for Case 1. This suggests that enlarging  $N$ is beneficial in scenarios where the inverse estimation problem is asymptotically linear or near-linear, but provides limited improvement under severe non-linear distortions.

\subsection{Open-Loop MIMO  Systems}
In this subsection, we investigate DeepJSCC-enabled image transmission over open-loop MIMO systems using ICL. To demonstrate the superiority of the proposed scheme, we consider several baseline approaches for comparison:
\begin{itemize}
	\item \textbf{Upper bound:} The channel is assumed to be perfectly known at the receiver. The LS technique is applied to recover the transmitted symbols from the noisy channel output. The perfect channel heatmap is also provided as side information to the DeepJSCC decoder during training.
	\item \textbf{Joint Design 1, CHM (ICAR):}  The DeepJSCC model and  ICL denoiser are jointly trained based on the loss function defined in  \eqref{trainingloss}.
	In addition, the DeepJSCC decoder is enhanced with additional context information derived from the channel. Two variants are considered under this design: one that employs a channel heatmap, and the other utilizes the proposed ICAR. Both variants correspond to the architecture shown in Fig.~\ref{Architecture_open_loop_icl}(b).
		\item \textbf{Joint Design 1, w/o CHM/ICAR:}  Similar to  ``Joint Design 1", but without channel heatmap and ICAR for the DeepJSCC decoder during training.
			
	\item \textbf{Joint Design 2,  CHM (ICAR):} This scheme is similar to ``Joint Design 1, CHM (ICAR)'', but  corresponds to the architecture shown in Fig.~\ref{Architecture_open_loop_icl}(a).
	\item \textbf{Joint Design 2, w/o CHM/ICAR:}  Similar to ``Joint Design 2", but without channel heatmap and ICAR.
	\item \textbf{Separate Design:} The DeepJSCC model and ICL denoiser are trained separately. Specifically, the ICL denoiser is first pre-trained and then integrated into the DeepJSCC framework. Subsequently, end-to-end training is performed for the DeepJSCC model while keeping the parameters of the ICL denoiser fixed. Note that the channel heatmap and ICAR are not used in this scheme.
	\item \textbf{LS Channel Equalization:}  The LS technique is first applied to estimate CSI based on pilot sequences and their outputs. Then, using the estimated CSI, the LS method is employed to recover the transmitted signal from the noisy channel output. No additional context information is provided to the DeepJSCC decoder.
	\item \textbf{BPG-Capacity:}  This approach represents a traditional separation-based method, where images are first compressed using the  BPG codec. The compressed bitstream is then encoded at a rate that matches the channel capacity.
\end{itemize}

\begin{figure}[!t]
	\centerline{\includegraphics[width=3.5in]{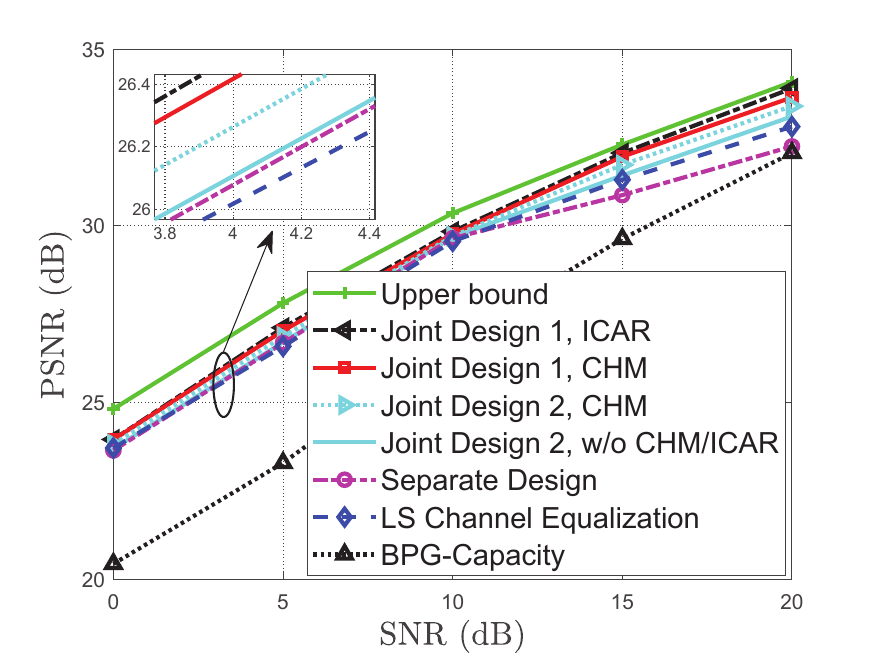}}
	\caption{PSNR versus SNR for open-loop MIMO systems under the perfect IQ case with $N=11$.} \label{Open_loop: perfectIQ}
%	\vspace{-0.3cm}
\end{figure}

In Fig.~\ref{Open_loop: perfectIQ}, we evaluate the PSNR performance versus SNR for open-loop MIMO systems under perfect IQ conditions. The results show that DeepJSCC-related schemes consistently outperform the conventional BPG-capacity scheme across all SNR values even though the latter is assumed to communicate at a rate that is equal to the capacity. We also observe that the separate design scheme slightly outperforms the LS channel equalization scheme at low SNR levels, although it becomes inferior at high SNR levels. This observation contrasts with the earlier finding in Fig.~\ref{MSEvsSNR}, where the ICL denoiser consistently achieved lower MSE than LS across all SNRs. This discrepancy is explained by the data distribution: in Fig.~\ref{MSEvsSNR}, both transmit pilots and queries are drawn from the same distribution during ICL training, whereas in DeepJSCC, the query data has a different distribution, leading to a performance degradation. Furthermore, when jointly training the ICL denoiser and the DeepJSCC model, we find that ``Joint Design 2, w/o CHM/ICAR" achieves higher PSNR across all SNR values, which aligns with the observations from Fig.~\ref{MSEvsSNR}. Additionally, incorporating channel context information, i.e., the channel heatmap or ICAR, into the DeepJSCC decoder input provides slight performance improvements. This is because context information enables the DeepJSCC decoder to better adapt its reconstruction process to the specific channel condition. Finally, it is observed that  ``Joint Design 1, CHM" scheme outperforms ``Joint Design 2, CHM" across all SNR values. This indicates that the ICL denoiser performs more effectively with a channel inversion model compared to a non-inverted channel model.

\begin{figure}[!t]
	\centerline{\includegraphics[width=3.5in]{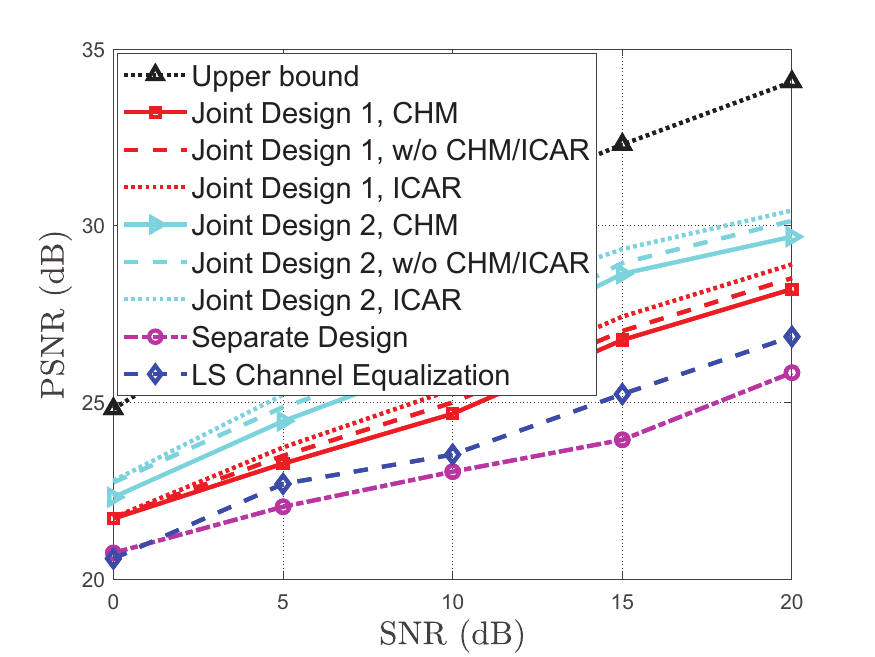}}
	\caption{PSNR versus SNR for open-loop MIMO systems  under  IQ imbalance Case 1 with $N=11$.} \label{Open_loop_IQI: case1}
%	\vspace{-0.3cm}
\end{figure}
In Fig.~\ref{Open_loop_IQI: case1}, we evaluate the PSNR performance versus SNR for open-loop MIMO systems under the IQ imbalance Case 1 scenario. It is observed that the PSNR achieved by all schemes, except for the upper bound scheme, significantly degrades compared to the perfect IQ case. This performance drop is due to the increased difficulty in learning the latent space associated with the inverse non-linear problem compared to the inverse linear case. Additionally, we observe that the ``Joint Design 1, CHM" scheme consistently underperforms compared to ``Joint Design 2, CHM" across all SNR levels. This can be attributed to the fact that the channel estimated by LS fails to accurately capture the true non-linear mapping function. Similarly, it is noted that incorporating additional channel heatmap  degrades system performance under severe IQ imbalance, contrary to the observations made in Fig.~\ref{Open_loop: perfectIQ} for the perfect IQ case. 
However,   it can be seen that  using ICAR yields slight improvements compared to their counterparts that use or omit the channel heatmap. This suggests that the DeepJSCC architecture is capable of learning the non-linear channel estimation directly from pilot sequences and their outputs.

\begin{figure}[!t]
	\centerline{\includegraphics[width=3.5in]{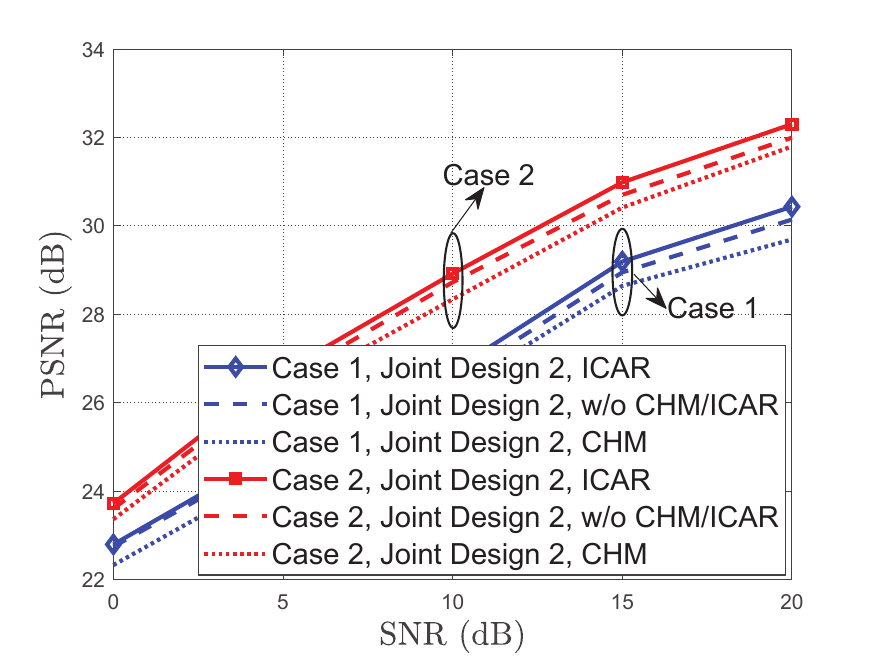}}
	\caption{Open-loop MIMO system in IQ imbalance Case 1 and  IQ imbalance Case 2   with $N=11$.} \label{Open_loop_IQI: case1vscase2}
%	\vspace{-0.3cm}
\end{figure}

In Fig.~\ref{Open_loop_IQI: case1vscase2}, we investigate the impact of different levels of IQ imbalance distortion on PSNR performance for open-loop MIMO systems with $N=11$. The results indicate that PSNR under IQ Case 1 is significantly lower than under IQ Case 2, demonstrating that more severe hardware impairments lead to greater system performance degradation. Moreover, schemes using channel heatmap as context information yield lower PSNR compared to those without channel heatmap, due to the channel estimation mismatch under severe distortion. Importantly, system performance can be improved by adopting ICAR, as demonstrated by the ``Joint Design 1, ICAR" and ``Joint Design 2, ICAR" schemes.

\begin{figure}[!t]
	\centerline{\includegraphics[width=3.5in]{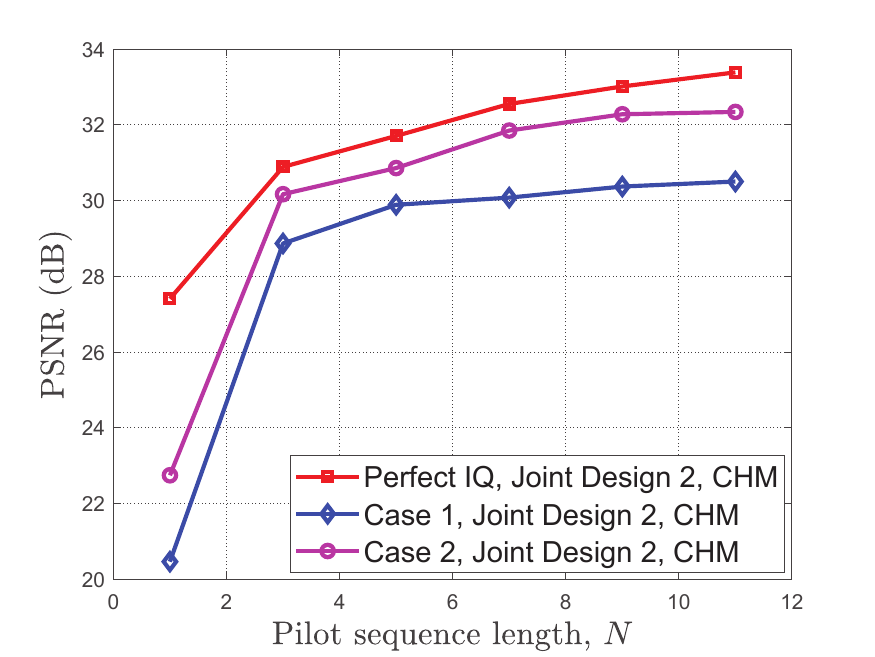}}
	\caption{PSNR versus pilot sequence length for open-loop MIMO systems  with ${\rm SNR}=20~$dB.} \label{Open_loop_VSsequencelength}
%	\vspace{-0.3cm}
\end{figure}

In Fig.~\ref{Open_loop_VSsequencelength}, we study the relationship between pilot sequence length and PSNR for open-loop MIMO systems under different IQ imbalance cases, with SNR fixed at ${\rm SNR}=20~$dB. For the perfect IQ case, the PSNR achieved by the ``Joint Design 2, CHM" scheme increases monotonically with $N$, indicating that utilizing pilot pairs as context information significantly enhances image reconstruction quality. For the IQ imbalance scenarios, we observe that the PSNR improves very slowly when $N$ exceeds $7$, and becomes nearly constant thereafter, particularly for IQ Case 1. This finding is consistent with the trends observed in Fig.~\ref{MSEvspilotlength} and Fig.~\ref{IQI_MSEVSpilotlength}, where the MSE achieved by the ICL denoiser under perfect IQ conditions decreases more rapidly than that under IQ imbalance conditions.

\subsection{Closed-Loop MIMO Systems}
In this subsection, we study DeepJSCC-enabled image transmission over closed-loop MIMO systems via ICL.  The benchmarks considered are similar to those used in the open-loop MIMO systems section, with two key differences: 1) Channel heatmap or ICAR is fed to both the DeepJSCC encoder and decoder when additional context information is considered; 2) The SVD is applied. 

\begin{figure}[!t]
	\centerline{\includegraphics[width=3.5in]{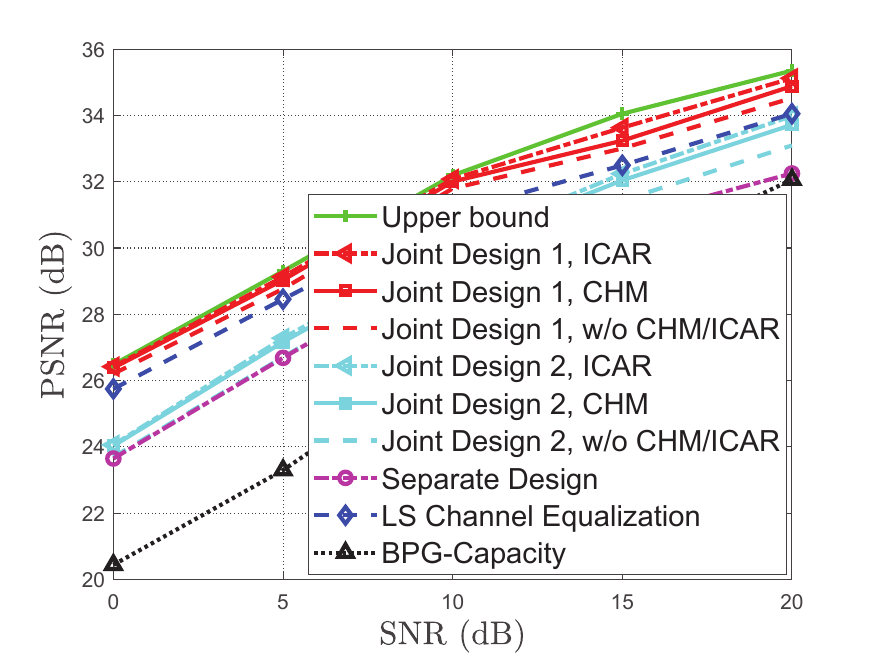}}
	\caption{PSNR versus SNR for closed-loop MIMO systems under the perfect IQ case with $N=11$.} \label{closed_loop: perfectIQ}
	\vspace{-0.3cm}
\end{figure}

In Fig.~\ref{closed_loop: perfectIQ}, we evaluate PSNR versus SNR for closed-loop MIMO systems under the perfect IQ condition. The results show that, except for the BPG-capacity and upper bound schemes, all other schemes achieve much higher PSNR compared to their counterparts in open-loop MIMO systems. This improvement is attributed to the fact that closed-loop DeepJSCC can actively shape the transmission to better match the channel conditions using SVD and adaptive encoding strategies.
Compared to the separate design, both joint design schemes exhibit significant PSNR improvements, highlighting the benefits of jointly training the DeepJSCC model and the ICL denoiser. Furthermore, it is observed that the proposed ``Joint Design 1, CHM" and ``Joint Design 1, ICAR" schemes approach the upper bound performance, particularly at low SNR levels. This implies that the ICL denoiser can effectively estimate the transmitted symbols by utilizing the eigenvalues and eigenvectors obtained from the estimated CSI.

\begin{figure}[!t]
	\centerline{\includegraphics[width=3.5in]{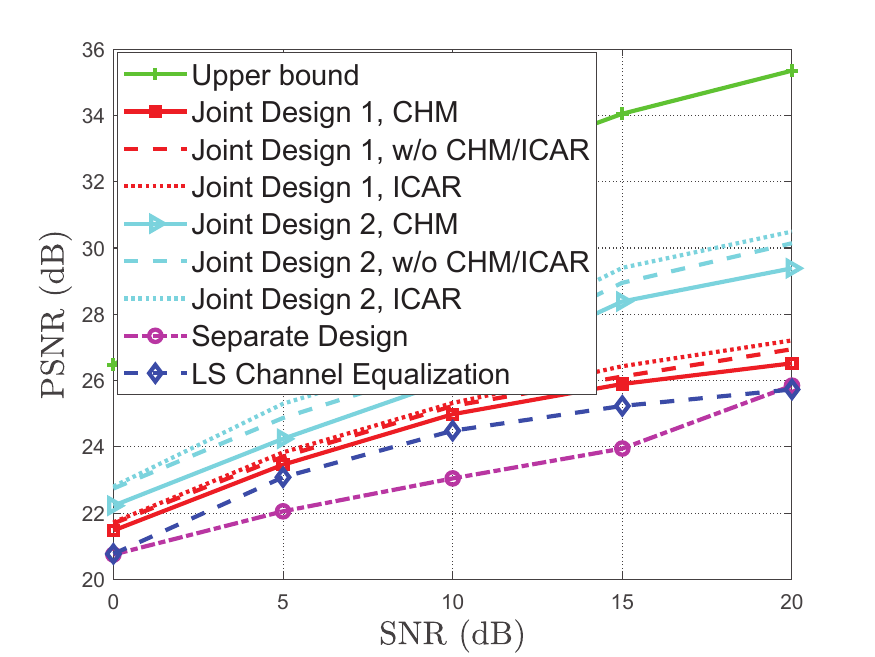}}
	\caption{PSNR versus SNR for closed-loop MIMO systems under IQ imbalance Case 1 with $N=11$. } \label{closed_loop_IQI: case1}
	\vspace{-0.3cm}
\end{figure}

In Fig.~\ref{closed_loop_IQI: case1}, we examine PSNR versus SNR for closed-loop MIMO systems under IQ imbalance Case 1. Similar to the observations in Fig.~\ref{Open_loop_IQI: case1}, the results show that incorporating channel heatmap as context information reduces the PSNR performance. This again confirms that, for inverse non-linear estimation problems, using CSI estimated by LS as context information degrades system performance.
Moreover, we observe that the separate design consistently performs the worst. However, for the joint design of the DeepJSCC model and the ICL denoiser, an acceptable PSNR can still be achieved when SNR exceeds $15~$dB. In addition, a slight performance improvement is observed when the ICAR is adopted.

\begin{figure}[!t]
	\centerline{\includegraphics[width=3.5in]{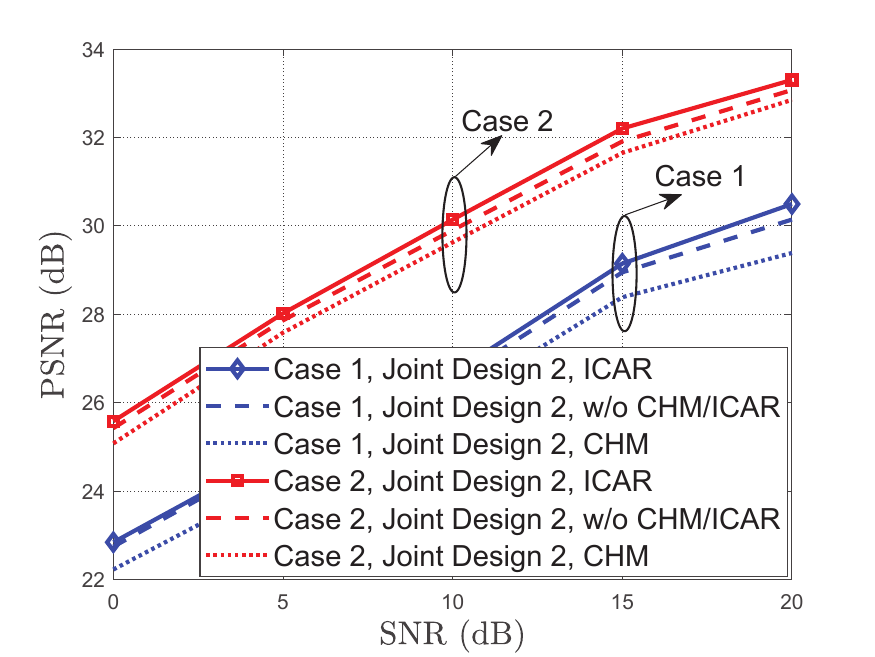}}
	\caption{IQ imbalance Case 1 versus IQ imbalance Case 2 for closed-loop MIMO systems with $N=11$. } \label{closed_loop_IQI: case1vscase2}
	\vspace{-0.3cm}
\end{figure}

In Fig.~\ref{closed_loop_IQI: case1vscase2}, we study the effect of IQ imbalance distortion on closed-loop MIMO systems with $N=11$. The PSNR achieved under IQ Case 1 is notably lower than that under IQ Case 2, indicating that hardware impairment significantly degrades symbol estimation performance. Additionally, the presence of channel heatmp as context information further harms the system in the presence of hardware impairments. These findings are consistent with those in Fig.~\ref{Open_loop_IQI: case1vscase2} for open-loop MIMO systems, reaffirming that hardware impairments are detrimental to inverse estimation in both open-loop and closed-loop settings. In addition, incorporating  ICAR  into both DeepJSCC encoder and decoder can improve image reconstruction quality.

\begin{figure}[!t]
	\centerline{\includegraphics[width=3.5in]{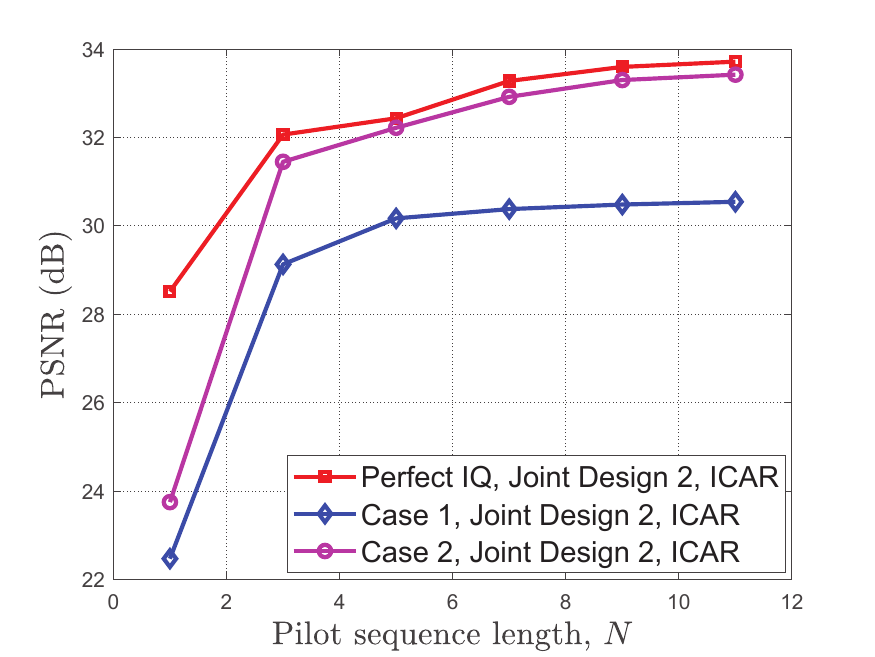}}
	\caption{PSNR versus pilot sequence length for closed-loop MIMO systems  with ${\rm SNR}=20~$dB.} \label{close_loop_VSsequencelength}
	\vspace{-0.3cm}
\end{figure}
In Fig.~\ref{close_loop_VSsequencelength}, we investigate the impact of pilot sequence length on PSNR for closed-loop MIMO systems under different IQ imbalance conditions at ${\rm SNR}=20~$dB. The results indicate that providing more context information through pilot sequences can significantly improve PSNR, leading to better image reconstruction quality. However, once $N\ge 7$, further increasing $N$ yields minimal performance gains for IQ imbalance cases, suggesting that a larger pilot sequence is unnecessary in such scenarios.
\begin{figure*}[!t]
	\centerline{\includegraphics[width=7in]{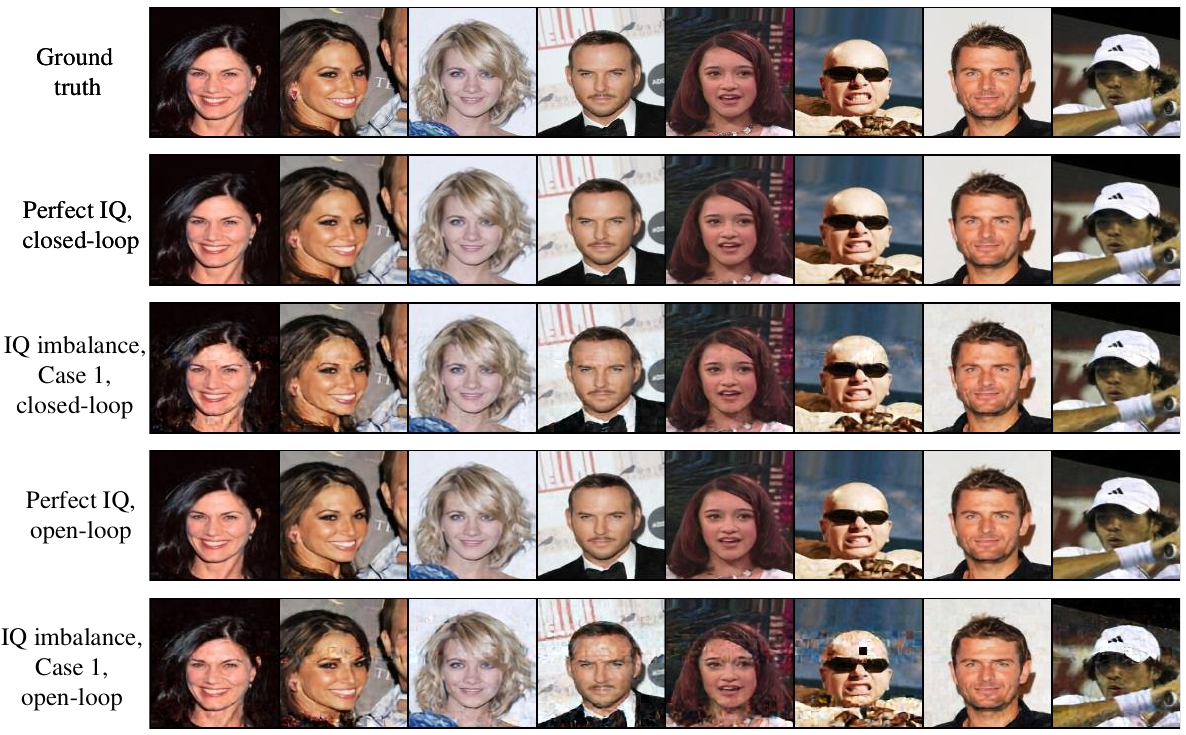}}
	\caption{Visualization of the reconstructed image from the CelebA dataset under ${\rm SNR}=10~$dB.} \label{vision_image}
	\vspace{-0.3cm}
\end{figure*}
\subsection{Visual Comparison}
Finally, a visual comparison of the reconstructed images under different schemes at ${\rm SNR}=10~$dB is presented in Fig.~\ref{vision_image}. We evaluate our proposed scheme using the CelebA dataset \cite{liu2015deep}, a large-scale facial attributes dataset containing over 200000 images of celebrities with 40 annotated attributes. As shown in Fig.~\ref{vision_image},  the top row displays the ground-truth images as reference. The subsequent rows illustrate the reconstructed images generated by different methods under various system impairments. Specifically, reconstructions are shown for: an SVD-based closed-loop system for perfect and imperfect IQ  under closed-loop, channel-inversion open-loop for perfect and imperfect IQ  under open-loop. It is observed that for perfect IQ balance, the transformer-based ICL for the DeepJSCC scheme effectively maintains high image fidelity across both closed-loop and open-loop systems. Even under challenging nonlinear channel impairments, the proposed scheme can still produce visually pleasing reconstructions. These results further demonstrate the effectiveness of the transformer-based ICL for DeepJSCC in MIMO image transmission.

\section{Conclusion}
In this work, we introduced a novel framework that integrates transformer-based ICL with DeepJSCC for efficient image transmission over MIMO channels. Both open-loop and closed-loop transmission scenarios were considered, including the realistic case of transceivers suffering from IQ imbalance. By employing the ICL denoiser that utilizes pilot sequences and their outputs as contextual information, our method enables end-to-end learning of encoding, decoding, and symbol estimation without explicit channel estimation. Extensive numerical evaluations validated the superior performance of the proposed approach over the traditional LS-based method and separate training strategy. In particular, the transformer-based ICL framework exhibited robust symbol estimation and improved image reconstruction quality even in the presence of severe hardware impairments.  Overall, our results suggest that transformer-based ICL offers a powerful paradigm shift for deep learning-based wireless communications, paving the way toward more resilient and efficient transmission systems for future 6G networks.

\bibliographystyle{IEEEtran}
\bibliography{Refdeepjscc.bib}

\end{document}